\documentclass[letterpaper,aps,prd,twocolumn,tightenlines,preprintnumbers,nofootinbib,showkeys,superscriptaddress,longbibliography
]{revtex4-1}
\pdfoutput=1
\usepackage{ragged2e}
\usepackage{comment}
\usepackage{XCharter}
\usepackage{mathptmx}
\usepackage{arydshln}
\usepackage[T1]{fontenc}
\usepackage{blkarray}
\usepackage{booktabs}
\usepackage{array}
\usepackage{fullpage}
\usepackage{amsfonts}
\usepackage{amsmath}
\usepackage{slashed}
\usepackage{amssymb}
\usepackage{graphicx}
\usepackage{epic}
\usepackage{eepic}
\usepackage{epsfig}
\usepackage{latexsym}
\usepackage[dvipsnames]{xcolor}
\usepackage[export]{adjustbox}
\usepackage{float}
\usepackage{multirow}
\usepackage{hyperref}
\usepackage{enumitem}
\usepackage{lipsum}
\hypersetup{colorlinks=true,citecolor=red,linkcolor=NavyBlue,urlcolor=NavyBlue}
\usepackage[caption=false,labelformat=simple]{subfig}
 
\usepackage{natbib}
\usepackage{relsize}
\usepackage[left=1.6cm,right=1.6cm,top=2.0cm,bottom=2.0cm]{geometry}
\usepackage{comment}
\usepackage{shorthand}
\usepackage{dcolumn}
\usepackage{siunitx}

\usepackage{pifont}
\begin{document}

\title{Observables and conformal properties of dark matter admixed isentropic neutron stars}

\author{Arijit Das}
\email{arijit21@iisertvm.ac.in}
\affiliation{Indian Institute of Science Education and Research Thiruvananthapuram, Vithura, Kerala, 695 551, India}

\author{Prashanth Jaikumar}
\email{prashanth.jaikumar@csulb.edu}
\affiliation{California State University Long Beach, Long Beach, California USA 90840}

\author{Adarsh Karekkat}
\email{adarsh.karekkat@unicaen.fr}
\affiliation{Université de Caen Normandie, ENSICAEN, CNRS/IN2P3, LPC Caen UMR6534, F-14000 Caen, France}
\affiliation{Indian Institute of Science Education and Research Thiruvananthapuram, Vithura, Kerala, 695 551, India}

\author{Tanumoy Mandal}
\email{tanumoy@iisertvm.ac.in}
\affiliation{Indian Institute of Science Education and Research Thiruvananthapuram, Vithura, Kerala, 695 551, India}

\begin{abstract}
\noindent
We construct an equation of state for isentropic dark-matter-admixed neutron stars (DMANS) with a hot core and relatively cold crust incorporating self-consistent temperature and DM density profiles for GeV-scale fermionic DM. 
We show that the enhancement of central stellar density due to DM accumulation, previously reported for cold neutron stars, remains robust. Substantial observable effects of DM accumulation arise only for sufficiently massive stellar configurations. Similar to earlier studies of cold NS, the speed of sound profile is shown to exhibit non-monotonic behavior for sufficiently large DM density in the core. We identify a competition between thermal effects due to nonzero values of entropy per baryon and softening effects of the dark sector which drive macroscopic properties and conformality indicators in opposite directions. This competition determines the onset of conformality near the stellar core and indicates that conformality signatures attributed to quark-matter in cold NS could be mimicked by DM admixture in isentropic stars.

\end{abstract}
\maketitle 


\section{Introduction}

Ultracompact objects such as neutron stars (NS) are a natural testing ground for an indirect probe of dark matter (DM) \cite{Bertone:2007ae}. Due to their extreme densities and strong gravitational fields, sufficient amounts of accumulated DM \cite{Narain:2006kx,Kouvaris:2010vv,Panotopoulos:2017pgv} may produce observable effects on its macroscopic properties  \cite{Guha:2021njn,Leung:2022wcf,Liu:2024rix,Koehn:2024gal} and internal dynamics such as rotational characteristics \cite{Issifu:2025gsq}, thermal evolution \cite{Kouvaris:2007ay,Baryakhtar:2017dbj,Keung:2020teb,AngelesPerez-Garcia:2022qzs,Alvarez:2023fjj,Avila:2023rzj,Bell:2023ysh,Zhou:2025dmy,Issifu:2025jac} and oscillation modes \cite{Nelson:2018xtr,Sotani:2025hzb,Shirke:2025lsu,Routaray:2025gbq,Sotani:2025lzy}. The highly unconstrained mass of DM ($10^{-22}~\text{eV} < M_\chi < 10^{19}~\text{GeV}$) already provides motivation for investigation of GeV-scale DM particles in NS cores. For instance, Ref.~\cite{Ivanytskyi:2019wxd} shows that the detection of a $2M_\odot$ NS in the most central region of the Milky Way Galaxy will impose an upper limit on the mass of dark matter particles of  $\sim 60$~GeV. As galaxies rotate, NS can sweep through the galactic DM halo and accumulate large amounts of GeV-scale DM particles within the core. Such captured DM particles will have effects on NS observables, which can be used to constrain the DM parameter space. 

Other mechanisms such as DM bremsstrahlung \cite{Mambrini:2022uol,SuperCDMS:2023sql} and gravitational slingshot mechanism \cite{Brayeur:2011yw,Takatsy:2025nnw} can cause a sufficient amount of DM to accumulate inside NS. Self-annihilating WIMP~\cite{Kouvaris:2007ay,Arcadi:2017kky} capture can manifest itself in merger events detected by future gravitational wave (GW) observatories~\cite{Grippa:2024ach,Ellis:2017jgp,Bezares:2019jcb}. On the other hand, non-self-annihilating DM, e.g. asymmetric DM~\cite{Kaplan:2009ag,Cohen:2010kn,McDermott:2011jp,Petraki:2013wwa,Zurek:2013wia} (which naturally prefers DM mass of a few GeVs) or mirror DM~\cite{Issifu:2024htq} influences macroscopic observables, oscillation frequencies and GW signatures of NS~\cite{Baldes:2017rcu,Ellis:2017jgp,Fornal:2022qim}. Ref.~\cite{Keung:2020teb} argues that in a neutron dark decay model, a NS can have a substantial DM population.

\par Since DM accumulated inside a NS is expected to be most concentrated at the stellar core and to decrease radially outward, a self-consistent DM density profile is essential for accurately modeling DMANS. On the other hand, simulations of proto-neutron star evolution~\cite{Nagakura:2019tmy,Pascal:2022qeg,Fiorillo:2023frv} typically impose a constant entropy per baryon condition. This leads to a temperature profile that increases with baryon density and provides a more realistic description of the NS interior than the assumption of a uniform temperature for DMANS, which is generally inconsistent with Tolman's law on thermal gradients in a gravitationally bound relativistic star~\cite{Santiago:2018kds}.

\par Combined effects of self-consistent DM and temperature profiles can affect the speed of sound ($C_S$) profile as well, which, in turn, influences the stiffness of the equation of state (EoS) and macroscopic properties. Although many constraints on $C^2_S$ are available in the literature~\cite{Moustakidis:2016sab,Tews:2018kmu,Leonhardt:2019fua,Margaritis:2020onf,Kojo:2020krb,Altiparmak:2022bke,Brandes:2022nxa} and non-monotonic profile of $C_S$ has been repeatedly suggested \cite{Altiparmak:2022bke,Bedaque:2014sqa,Chatterjee:2023ecc,Mroczek:2023zxo,Chen:2023hqm,Yao:2023yda,Marczenko:2024uit,Das:2025skn} for cold NS, apart from causality bound $0 < C_S < 1$, almost no other such constraints exist for hot compact objects. Given such uncertainties, it would be instructive to study the effect of self-consistent DM and temperature profiles on the $C^2_S$. In addition, it would be interesting to study conformal properties in warm DMANS matter and compare profiles of conformality indicators, measures that indicate onset of deconfined quark matter \cite{Mandal:2009uk,Mandal:2012fq,Mandal:2016dzg,Mandal:2017ihr}, with conformality thresholds \cite{Annala:2019puf,Fujimoto:2022ohj,Annala:2023cwx,Marczenko:2023txe}.
\par In the present work, we study the effect of self-consistent temperature and GeV-scale DM density profiles on macroscopic observables and conformal behavior of isentropic DMANS under charge neutrality and $\beta$-equilibrium conditions. However, we do not assume that the DM distribution is in equilibrium with the hadronic/leptonic sector, and hence, we consider the DM chemical potential $\mu_\chi$ as a free parameter. The paper is structured in the following way: In Section~\ref{model}, we discuss the model, hadronic and dark sector parameter sets, nucleon effective mass and temperature profiles. Section~\ref{Equation_of_State} discusses the EoS. In Section~\ref{Result}, we present numerical results for static and slowly rotating NSs for varying DM chemical potential. We discuss the effects on the speed of sound profile and its non-monotonic behavior in Section~\ref{SOS_Profile}. Section~\ref{radial_variations} presents results for radial variations of conformality indicators for varying stellar masses. Finally, we present our concluding remarks in Section~\ref{END}.

\begin{figure*}
    \centering
    \begin{minipage}{0.49\textwidth}
        \centering
        \includegraphics[width=\linewidth]{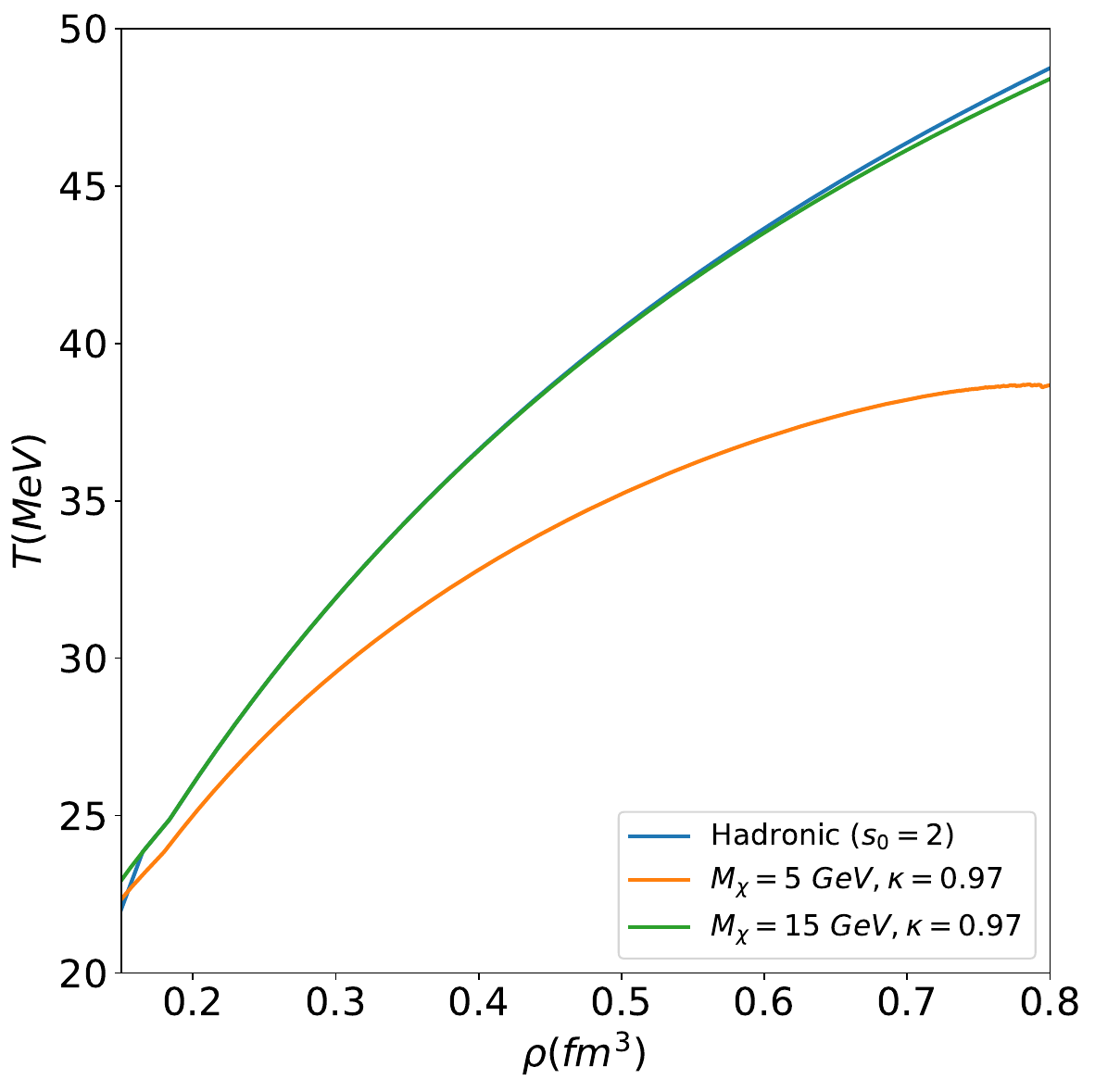}
        
        {\small (a)}
    \end{minipage}
    \hfill
    \begin{minipage}{0.49\textwidth}
        \centering
        \includegraphics[width=\linewidth]{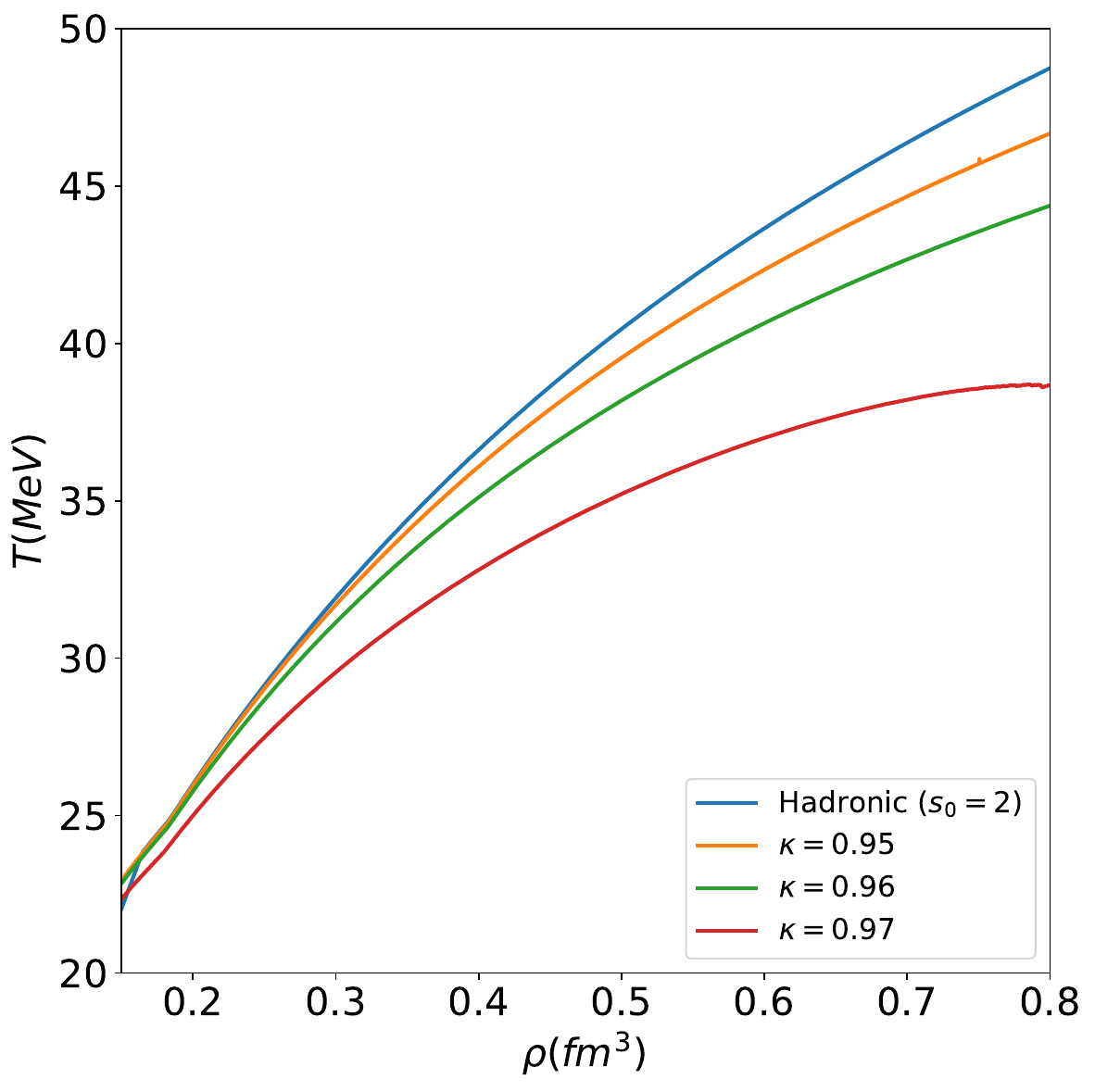}
        
        {\small (b)}
    \end{minipage}
    \caption{\justifying
    (a) Temperature profiles $T(\rho^{V}_{\text{tot}})$ as a function of the total baryon density $\rho^{V}_{\text{tot}}$ for $s_0=2$, comparing $M_{\chi}=5~\text{GeV}$ and $M_{\chi}=15~\text{GeV}$ and for fixed $\kappa = 0.97$. As  $s_0 = 0$ corresponds to the profile $T(\rho^V_{\text{tot}}) = 0$, its explicit profile has not been shown in the figure. Whereas increasing $s_0$ amplifies the temperature profile, the presence of DM dampens it, indicating a competition between thermal enhancement due to increasing $s_0$ and the cooling induced by DM admixture. For fixed $M_\chi$ and $\kappa$, lighter DM particles lead to a cooler stellar core. (b) Temperature profiles $T(\rho^{V}_{\text{tot}})$ for $s_0=2$ and $M_\chi=5~\text{GeV}$ for different values of $\kappa$. Increasing $\kappa$ results in a reduction of the NS temperature.}
    \label{fig:T_vs_Rho_Combined}
    \makeatletter
    \renewcommand{\@currentlabel}{\thefigure (a)}\label{fig:T_vs_Rho_a}
    \renewcommand{\@currentlabel}{\thefigure (b)}\label{fig:T_vs_Rho_b}
    \makeatother
\end{figure*}

\section{Model Selection and Parameter Sets}
\label{model}

\noindent In this work, we adopt the DM admixed $\mathrm{SU}(2)$ linear sigma model. Following Ref.~\cite{Guha:2021njn}, we present the relevant details of the model below.

\subsection{Model Lagrangian}
\label{Model_Selection}
\noindent
Apart from the kinetic contribution from the mesons, the hadronic part of the model is given as:
\begin{align}
    \mathcal{L}_{H} &\supset \overline{\psi}\slashed{D}\psi + \frac{1}{2}g_{\omega}^2x^2 \omega_{\mu}\omega^{\mu} + \frac{1}{2}m^2_{\rho}\vec{\rho}_{\mu}\cdot\vec{\rho}^{\mu} -U(\bar{x}).
\end{align}
Here, $\psi$ is the $\mathrm{SU}(2)$ nucleon isospin doublet. The operator $\slashed{D}$ is defined as,
\begin{align}
    \slashed{D} &= i\gamma_{\mu}\partial^{\mu} - g_{\sigma}(\sigma + i\gamma_{5}\vec{\tau}\cdot\vec{\pi})
    - g_{\omega}\gamma_{\mu}\omega^{\mu} 
    - \frac{1}{2}g_{\rho}\gamma_{\mu}\vec{\tau}\cdot\vec{\rho}^{\mu}, 
\end{align}
The scalar potential $U(\bar{x})$ is defined as
\begin{align}
\label{eq:sclrpot}
    U(\bar{x}) = \frac{\lambda}{4}\bar{x}^{4} + \frac{B}{6}\bar{x}^{6} + \frac{C}{8}\bar{x}^{8},
\end{align}
where $\bar{x}^2 = x^2 - x^2_0$ with $x^2 = \sigma^2 + \vec{\pi}^2$, $x_0$ being the vacuum expectation value of $\sigma$. The fermionic dark sector Lagrangian is given by
\begin{align}
    \mathcal{L}_{DM} = \overline{\chi}\left( i\gamma_{\mu}\partial^{\mu} - m_{\chi} \right)\chi,
\end{align}
Here, $\chi$ is the DM spinor and $m_{\chi}$ is its mass. The new scalar mediator $\phi$ and the new vector mediator $\xi$ connect the hadronic sector and the dark sector via the following interactions,
\begin{align}
    \mathcal{L}_{I} = - g_{\phi}\overline{\psi}\psi\phi - g_{\xi}\overline{\psi}\gamma_{\mu}\psi \xi^{\mu} -y_{\phi}\overline{\chi}\chi\phi -y_{\xi}\overline{\chi}\gamma_{\mu}\chi \xi^{\mu}.
\end{align}
The mass terms for $\phi$ and $\xi$ are given by
\begin{align}
    \mathcal{L}_{M} \supset \frac{1}{2}m^{2}_{\xi}\xi_{\mu}\xi^{\mu} - \frac{1}{2}m^2_{\phi}\phi^2,
\end{align}
The leptonic sector is completely decoupled from other sectors of the model and given by 
\begin{align}
    \mathcal{L}_{\ell} = \sum_{\ell}\overline{\ell}\left( i\gamma_{\mu}\partial^{\mu} - m_{\ell} \right)\ell.
\end{align}
Here, $\ell$ is the spinor field for the charge lepton of flavor $\ell$ of mass $m_{\ell}$. Collecting all the terms, the total model is given by
\begin{align}
    &\mathcal{L} \supset \mathcal{L}_{H} + \mathcal{L}_{DM} + \mathcal{L}_{I} + \mathcal{L}_{M} + \mathcal{L}_{\ell}.
\end{align}
The masses of the nucleons, $\omega$ and $\sigma$ are denoted by $m_N$, $m_{\omega}$, $m_{\sigma}$, respectively. These masses arise from chiral symmetry breaking in the scalar sector of both the hadronic and dark sectors. If $\phi_0$ denotes the vacuum expectation values of the scalar field $\phi$, then the masses are given by,
\begin{align}
m_N = g_{\sigma}x_{0} + g_{\phi}\phi_0,\quad
m_{\omega} = g_{\omega}x_{0},\quad
m_{\sigma} = \sqrt{2\lambda}x_0.
\end{align}

\begin{figure*}[hbt]
    \centering

    \begin{minipage}[t]{0.48\textwidth}
        \centering
        \includegraphics[width=\textwidth]{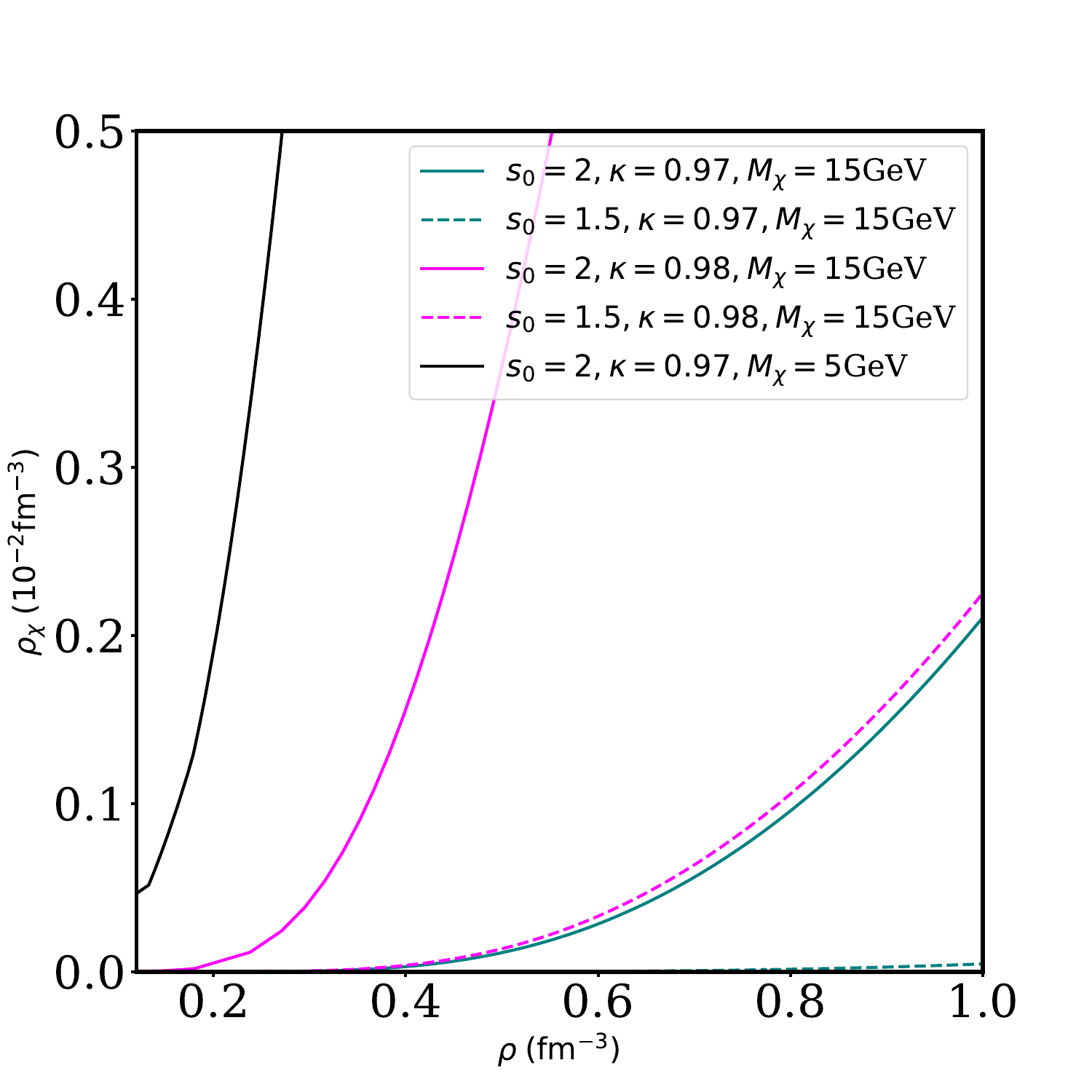}
        \par\smallskip
        (a)
    \end{minipage}
    \hfill
    \begin{minipage}[t]{0.48\textwidth}
        \centering
        \includegraphics[width=\linewidth]{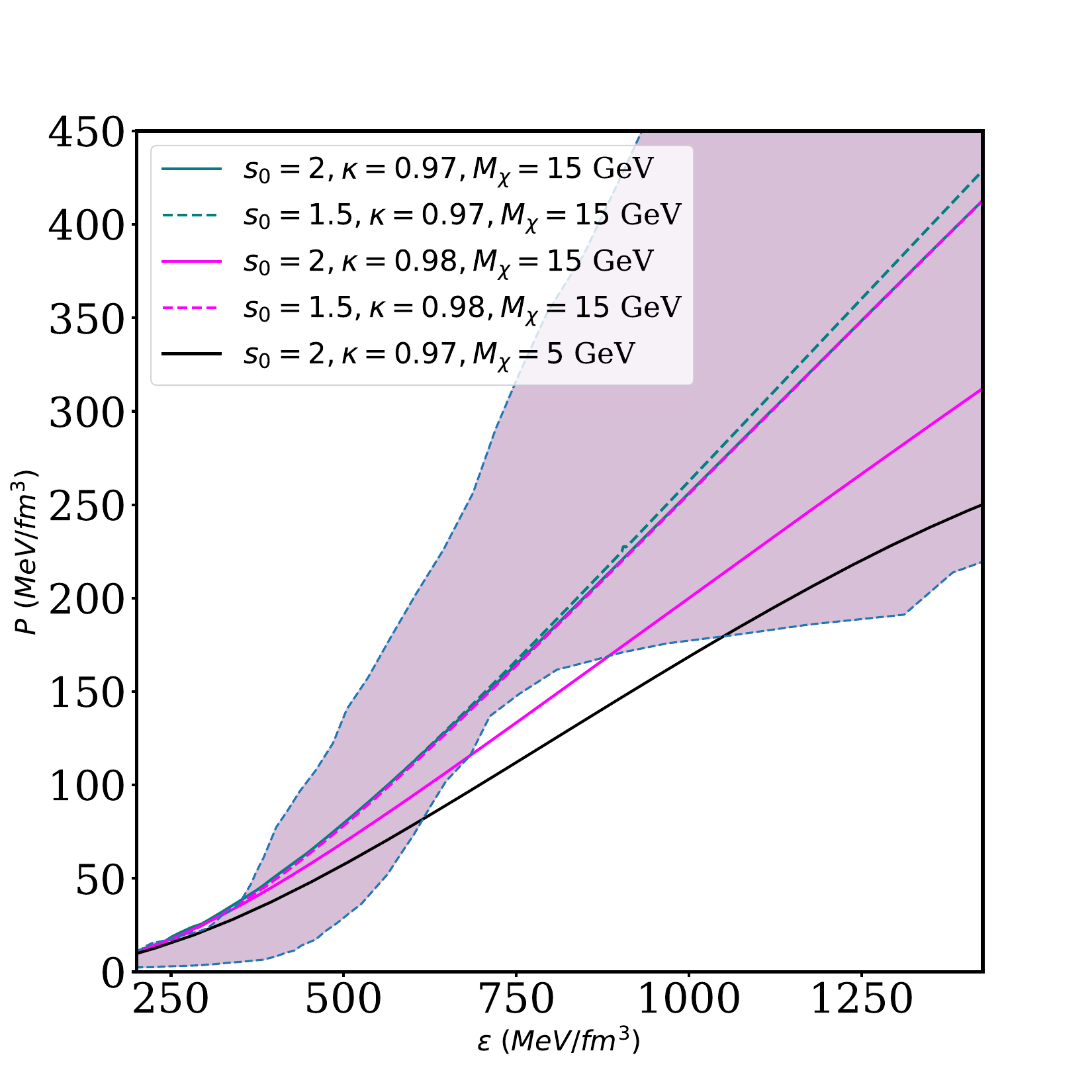}
        \par\smallskip
        (b)
        \label{fig:eos_spb_a}
    \end{minipage}

    \caption{\justifying
    (a) DM density profiles and (b) the corresponding EoSs for the same benchmarks. Results are shown for $\kappa = 0.97$ and $\kappa = 0.98$. The shaded region denotes the region $\Lambda(1.4M_\odot) < 400$ constraint as obtained from \cite{Annala:2017llu}. Comparing these two figures together, it is evident that more amplified DM density profiles lead to softer EoSs for sufficiently large energy densities.}
    \label{fig:dm_eos_combined}
    \makeatletter
    \renewcommand{\@currentlabel}{\thefigure (a)}\label{fig:dm_eos_combined_a}
    \renewcommand{\@currentlabel}{\thefigure (b)}\label{fig:dm_eos_combined_b}
    \makeatother
\end{figure*}

\begin{table*}[!htbp]
\centering
\renewcommand{\arraystretch}{1.5}
\footnotesize
\setlength{\tabcolsep}{4pt} 
\resizebox{\textwidth}{!}{
\begin{tabular}{llccccccccccccc}
\toprule
\toprule
\multicolumn{15}{c}{\textbf{Hadronic sector}} \\
\midrule
\midrule
& Parameter & $C_{\sigma}$ & $C_{\omega}$ & $C_{\rho}$ & $B$ & $C$ & $m_{\sigma}$ & $f_{\pi} = x_0$ & $K$ & $B.E./A$ & $J$ & $L_0$ & $\rho_0$ & $Y(\rho_0)$ \\
& Units      & (fm$^2$)     & (fm$^2$)     & (fm$^2$)   & (fm$^2$) & (fm$^4$) & (MeV) & (MeV) & (MeV) & (MeV) & (MeV) & (MeV) & (fm$^{-3}$) & -- \\
& Value     & 7.325        & 1.642        & 5.324      & $-$6.586 & 0.571    & 444.614 & 153.984 & 231 & $-$16.3 & 32 & 88 & 0.153 & 0.87 \\
\bottomrule
\bottomrule
\end{tabular}%
}
\caption{Parameter sets for hadronic sector \cite{Guha:2021njn,Jha:2008yth}.}
\label{tab:had_par}
\end{table*}

\vspace{0.5em}

\begin{table*}[!htbp]
\centering
\renewcommand{\arraystretch}{1.4}
\scriptsize
\setlength{\tabcolsep}{2.5pt} 
\resizebox{0.52\textwidth}{!}{
\begin{tabular}{llccccccccc}
\toprule
\toprule
\multicolumn{11}{c}{\textbf{Dark matter sector}} \\
\midrule
\midrule
& Parameter & \multicolumn{2}{c}{$m_{\chi}$ (GeV)} &  \multicolumn{2}{c}{$m_{\phi}$ (MeV)} & \multicolumn{2}{c}{$m_{\xi}$ (MeV)} & \multicolumn{2}{c}{$y_{\phi} = y_{\xi}$} & $g_{\phi} = g_{\xi}$ \\
& BP I   & \multicolumn{2}{c}{5}   & \multicolumn{2}{c}{9}    & \multicolumn{2}{c}{11}   & \multicolumn{2}{c}{0.13} & $1.1 \times 10^{-4}$ \\
& BP II  & \multicolumn{2}{c}{15}  & \multicolumn{2}{c}{20}   & \multicolumn{2}{c}{34}   & \multicolumn{2}{c}{0.21} & $1.1 \times 10^{-4}$ \\
\bottomrule
\bottomrule
\end{tabular}%
}
\caption{Parameter sets for DM sector \cite{Guha:2021njn,Sen:2021wev}.}
\label{tab:dark_par}
\end{table*}

\begin{figure*}[t]
    \centering

    \begin{minipage}{0.48\textwidth}
        \centering
        \includegraphics[width=\linewidth]{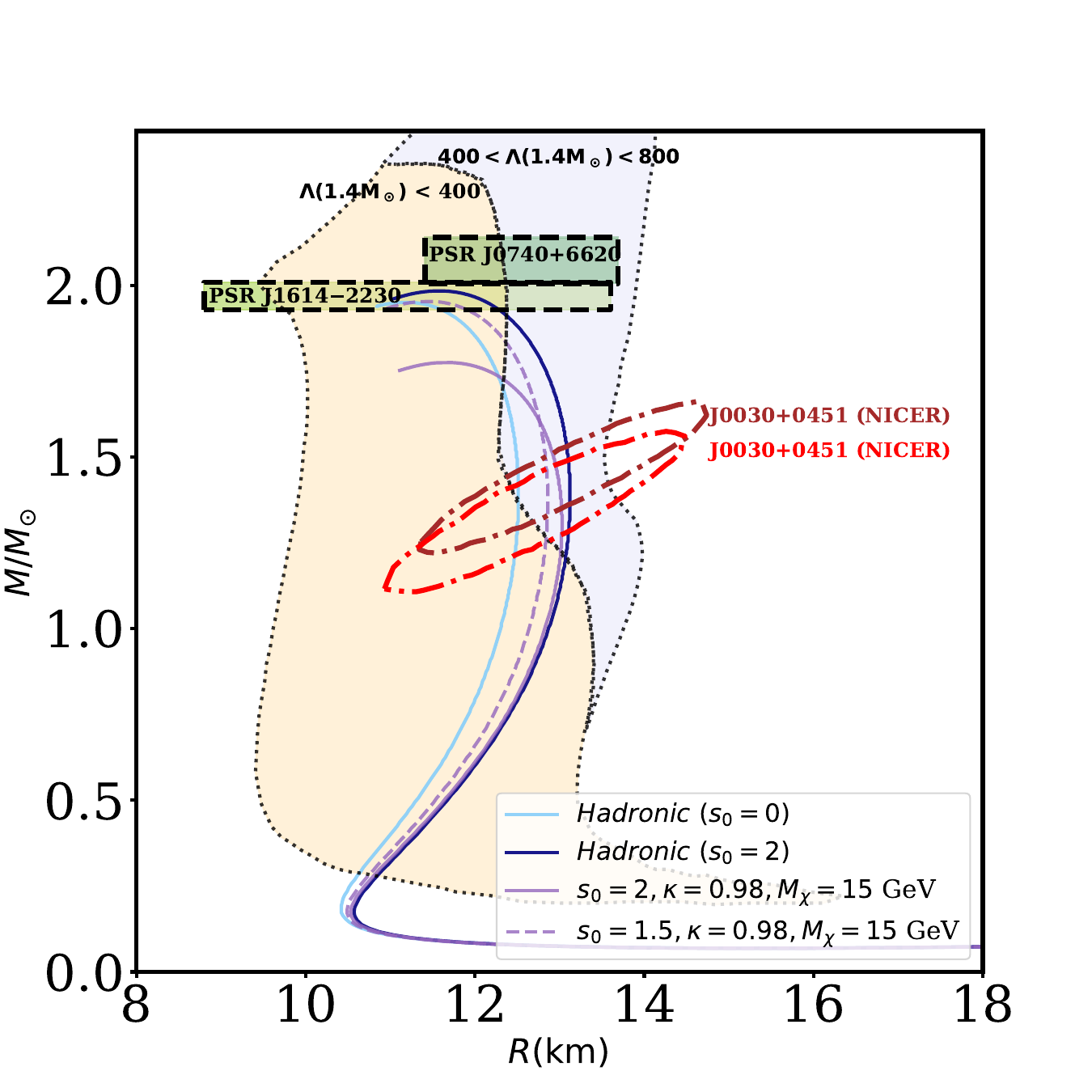}
        \par\smallskip
        (a)
    \end{minipage}
    \hfill
    \begin{minipage}{0.48\textwidth}
        \centering
        \includegraphics[width=\linewidth]{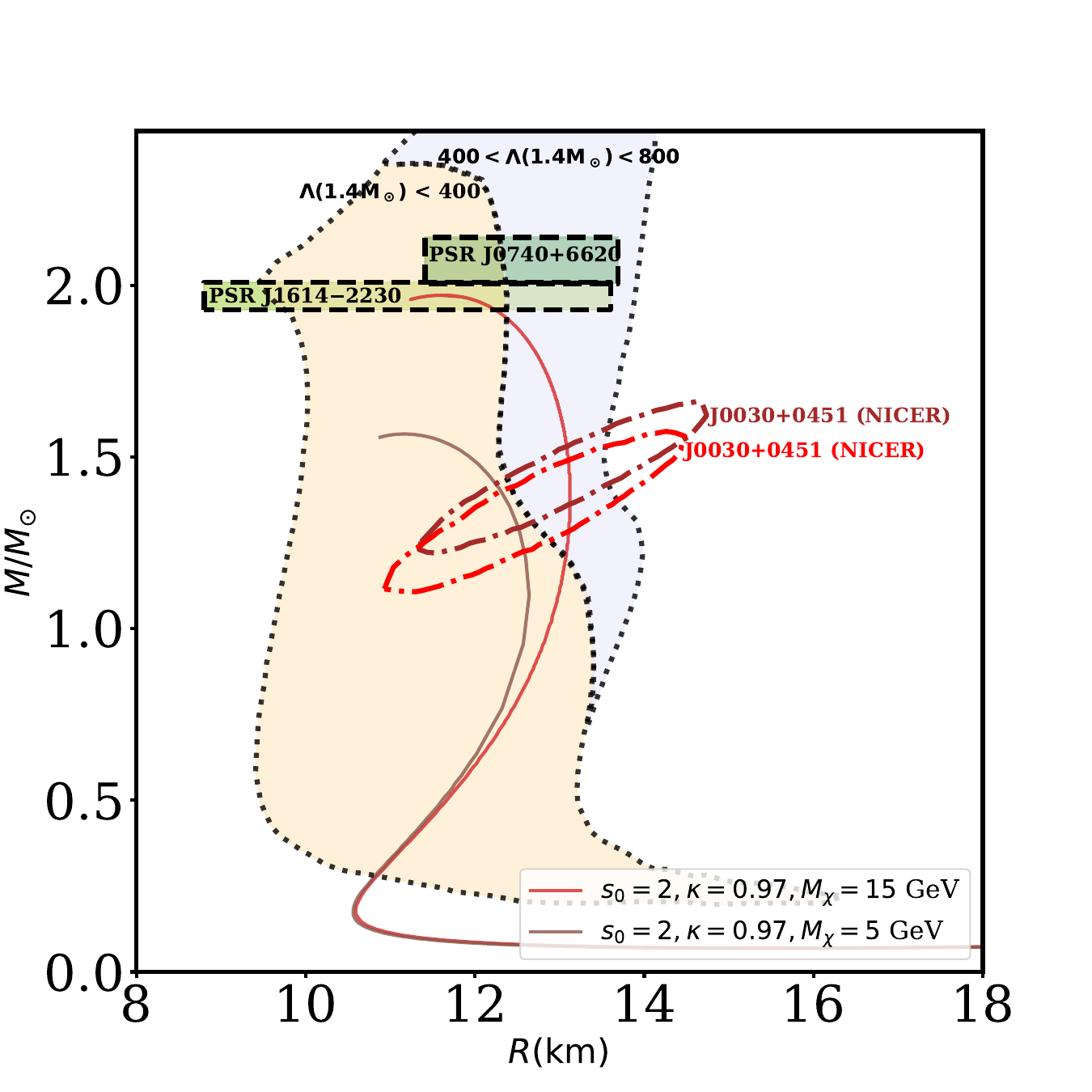}
        \par\smallskip
        (b)
    \end{minipage}

    \caption{\justifying
    $M$--$R$ relations. (a) Competing mechanism between thermal effects and the dark sector for $\kappa = 0.98$ and $M_\chi = 15~$GeV and effect on maximum mass for varying $s_0$. Whereas increasing $s_0$ from $s_0 = 0$ to $s_0 = 2$ increases the mass, inclusion of DM further decreases it, showing a competition between the thermal effects and dark sector. (b) The reduction of maximum mass for lighter DM particles for $s_0 = 2$ and $\kappa = 0.97$.}
    \label{fig:RxMy_2x4}
    \makeatletter
    \renewcommand{\@currentlabel}{\thefigure (a)}\label{fig:RxMy_2x4_a}
    \renewcommand{\@currentlabel}{\thefigure (b)}\label{fig:RxMy_2x4_b}
    \makeatother
\end{figure*}

\begin{table*}[hbt]
\centering
\begin{tabular}{c c cc ccc}
\toprule
& \textbf{Hadronic ($s_0 = 2$)}
& \multicolumn{2}{c}{\textbf{BPI} ($s_0=2$)}
& \multicolumn{3}{c}{\textbf{BPII} ($s_0=2$)} \\
\cmidrule(lr){2-2}
\cmidrule(lr){3-4}
\cmidrule(lr){5-7}
\textbf{Quantity}
&
& 0.96 & 0.97
& 0.96 & 0.97 & 0.98 \\
\midrule
$R_{\max}\;(\mathrm{km})$
& 11.58
& 11.45 & 11.14
& 11.58 & 11.62 & 11.67 \\
$M_{\max}\;(M_\odot)$
& 1.98
& 1.82 & 1.57
& 1.98 & 1.97 & 1.77 \\
$\Lambda_{1.4}$
& 878.35
& 785.31 & 511.586
& 886.44 & 881.82 & 814.85 \\
\bottomrule
\end{tabular}
\caption{\justifying Comparison of maximum mass ($M_{\text{max}}$), corresponding radius ($R_{\text{max}}$), and $\Lambda_{\text{1.4}}$ for BPI and BPII parameter sets at $s_0=2$ for different values of $\kappa$.}
\label{tab:BP1_BP2_mixedalpha}
\end{table*}

\begin{figure*}[t]
    \centering

    \begin{minipage}{0.48\textwidth}
        \centering
        \includegraphics[width=\linewidth]{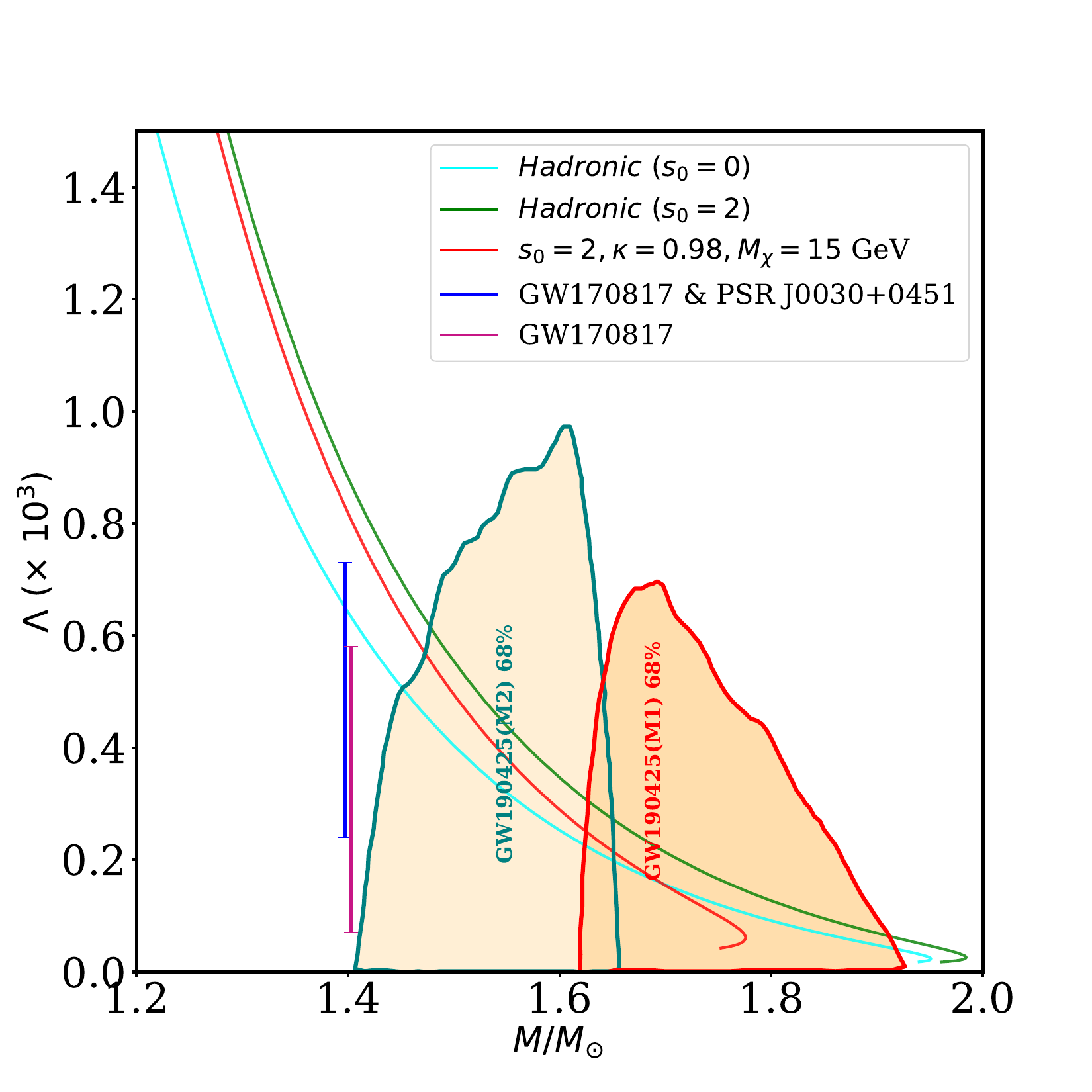}
        \par\smallskip
        (a)
    \end{minipage}
    \hfill
    \begin{minipage}{0.48\textwidth}
        \centering
        \includegraphics[width=\linewidth]{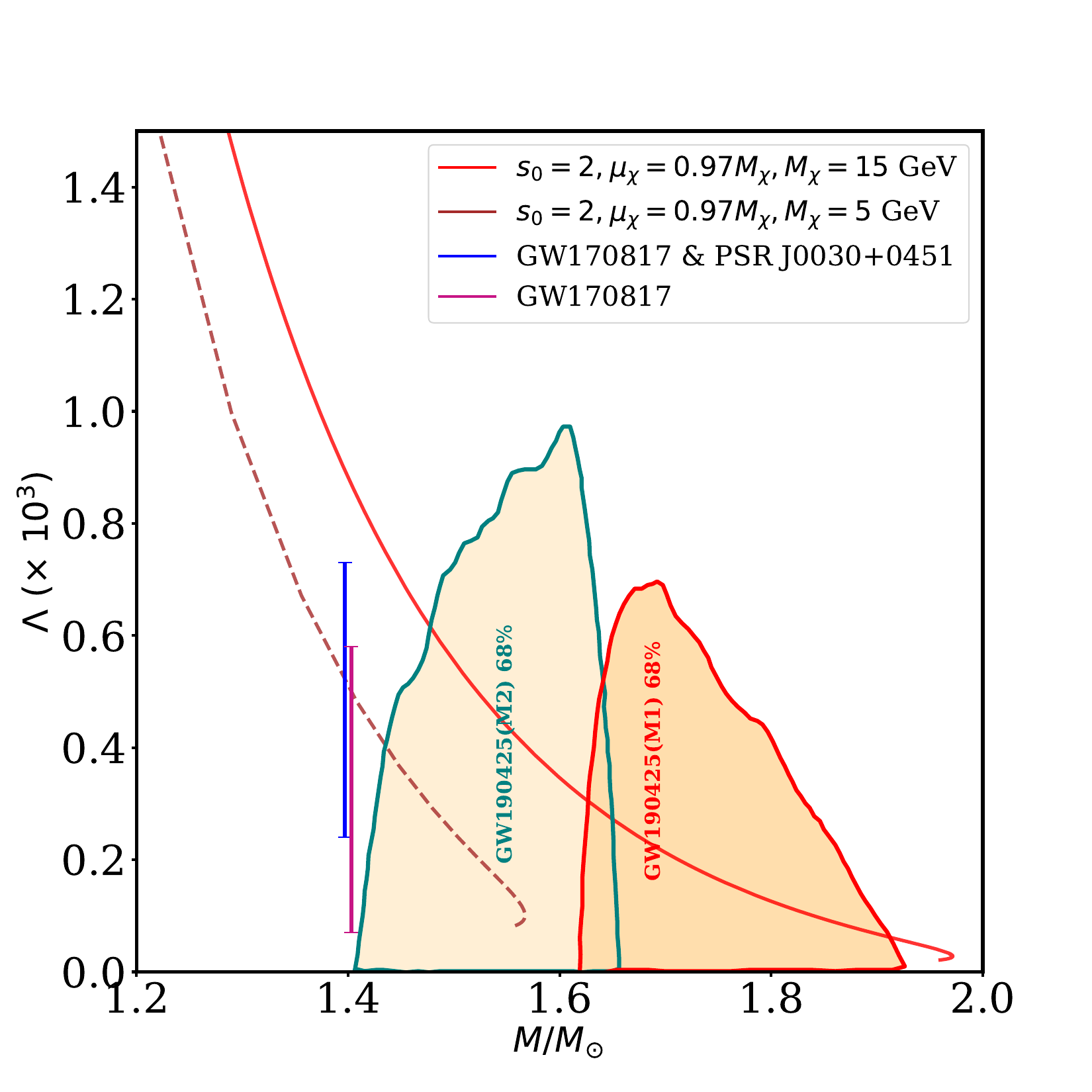}
        \par\smallskip
        (b)
    \end{minipage}

    \caption{\justifying
    $M$--$\Lambda$ relations. (a) The competing mechanism between thermal effects and the dark sector for $\kappa = 0.98$ and $M_\chi = 15~$GeV and effect on $\Lambda_{\text{1.4}}$ for varying $s_0$. Whereas increasing $s_0$ from $s_0 = 0$ to $s_0 = 2$ increases $\Lambda_{\text{1.4}}$, including DM further decreases it. (b) The reduction of $\Lambda_{\text{1.4}}$ for lighter DM particles for $s_0 = 2$ and $\kappa = 0.97$.}
    \label{fig:M_Lambda_2x2}
    \makeatletter
    \renewcommand{\@currentlabel}{\thefigure (a)}\label{fig:M_Lambda_2x2_a}
    \renewcommand{\@currentlabel}{\thefigure (b)}\label{fig:M_Lambda_2x2_b}
    \makeatother
\end{figure*}

\begin{figure*}[t]
    \centering
    \begin{minipage}{0.48\textwidth}
        \centering
        \includegraphics[width=\linewidth]{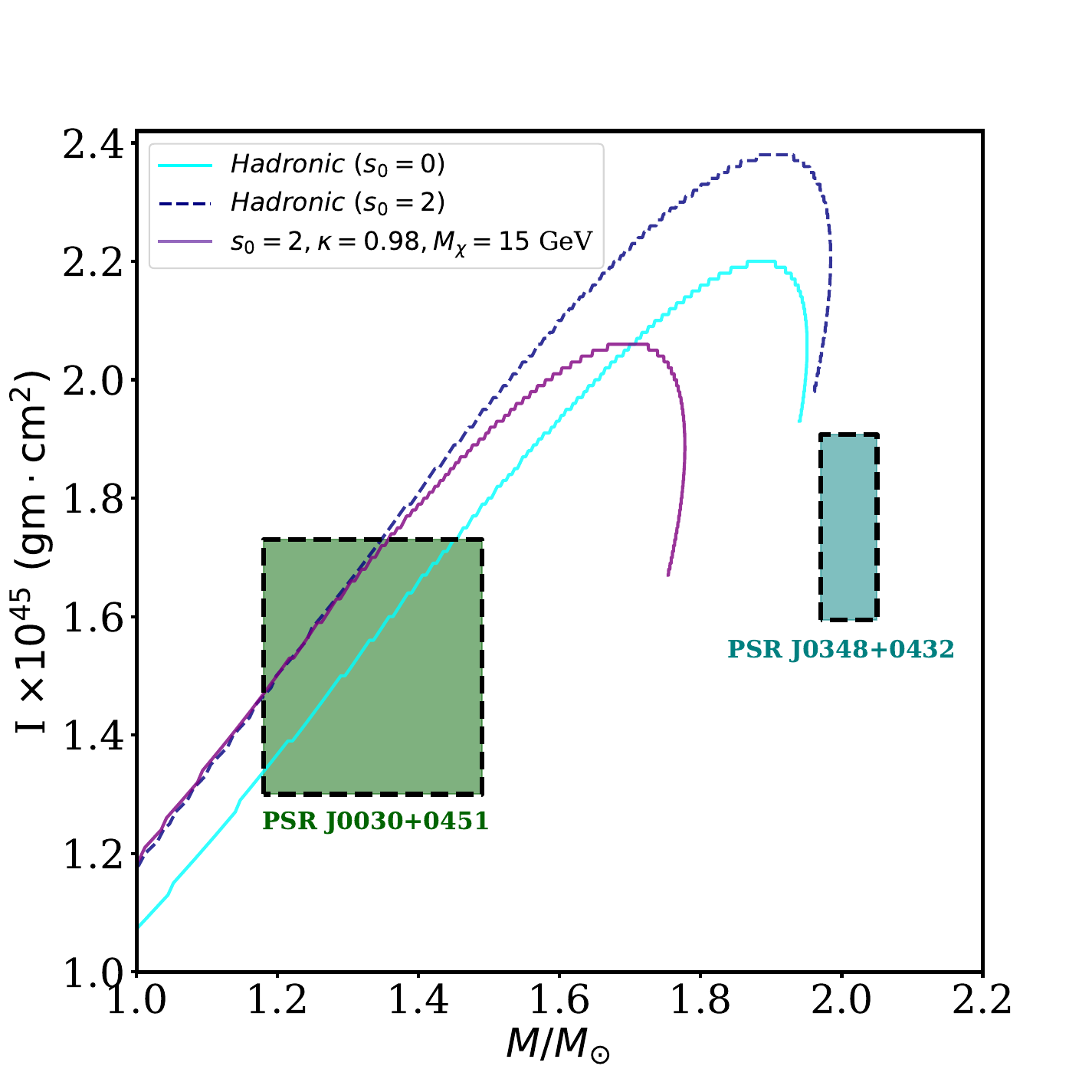}
        \par\smallskip
        (a)
    \end{minipage}
    \hfill
    \begin{minipage}{0.48\textwidth}
        \centering
        \includegraphics[width=\linewidth]{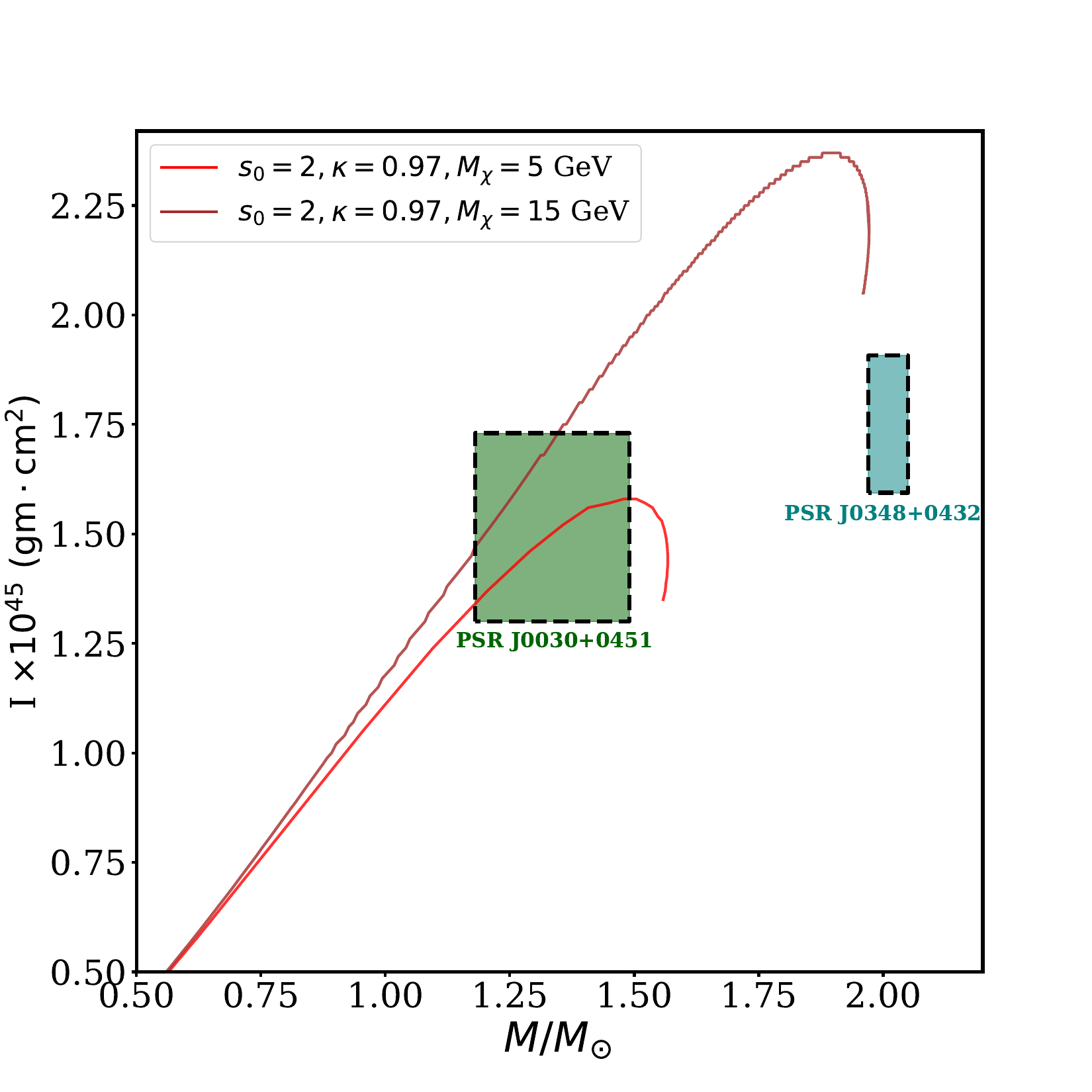}
        \par\smallskip
        (b)
    \end{minipage}

    \caption{\justifying
    $M$--$I$ relations. (a) The competing mechanism between thermal effects and the dark sector for $\kappa = 0.98$ and $M_\chi = 15~$GeV and effect on $I$ for varying $s_0$. Whereas increasing $s_0$ from $s_0 = 0$ to $s_0 = 2$ increases the MoI, for sufficiently heavy stellar configurations, inclusion of DM again decreases it. (b) The reduction of MoI at any stellar mass and also reduction of maximum MoI for lighter DM particles for $s_0 = 2$ and $\kappa = 0.97$.}
    \label{fig:MI_RI_combined}
    \makeatletter
    \renewcommand{\@currentlabel}{\thefigure (a)}\label{fig:MI_RI_combined_a}
    \renewcommand{\@currentlabel}{\thefigure (b)}\label{fig:MI_RI_combined_b}
    \makeatother
\end{figure*}

\begin{figure*}[t]
    \centering

    \begin{minipage}{0.48\textwidth}
        \centering
        \includegraphics[width=\linewidth]{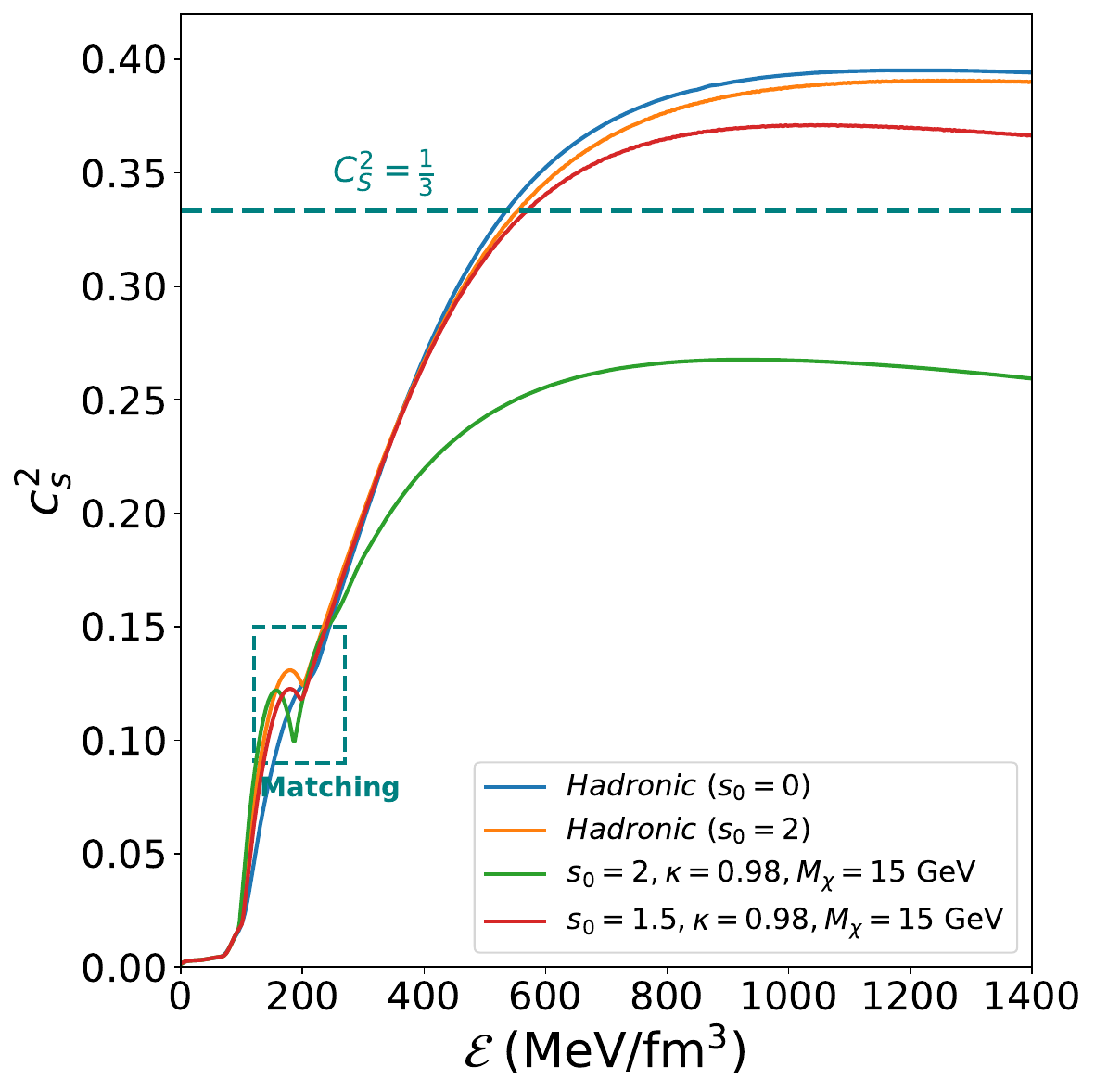}
        \par\smallskip
        (a)
    \end{minipage}
    \hfill
    \begin{minipage}{0.48\textwidth}
        \centering
        \includegraphics[width=\linewidth]{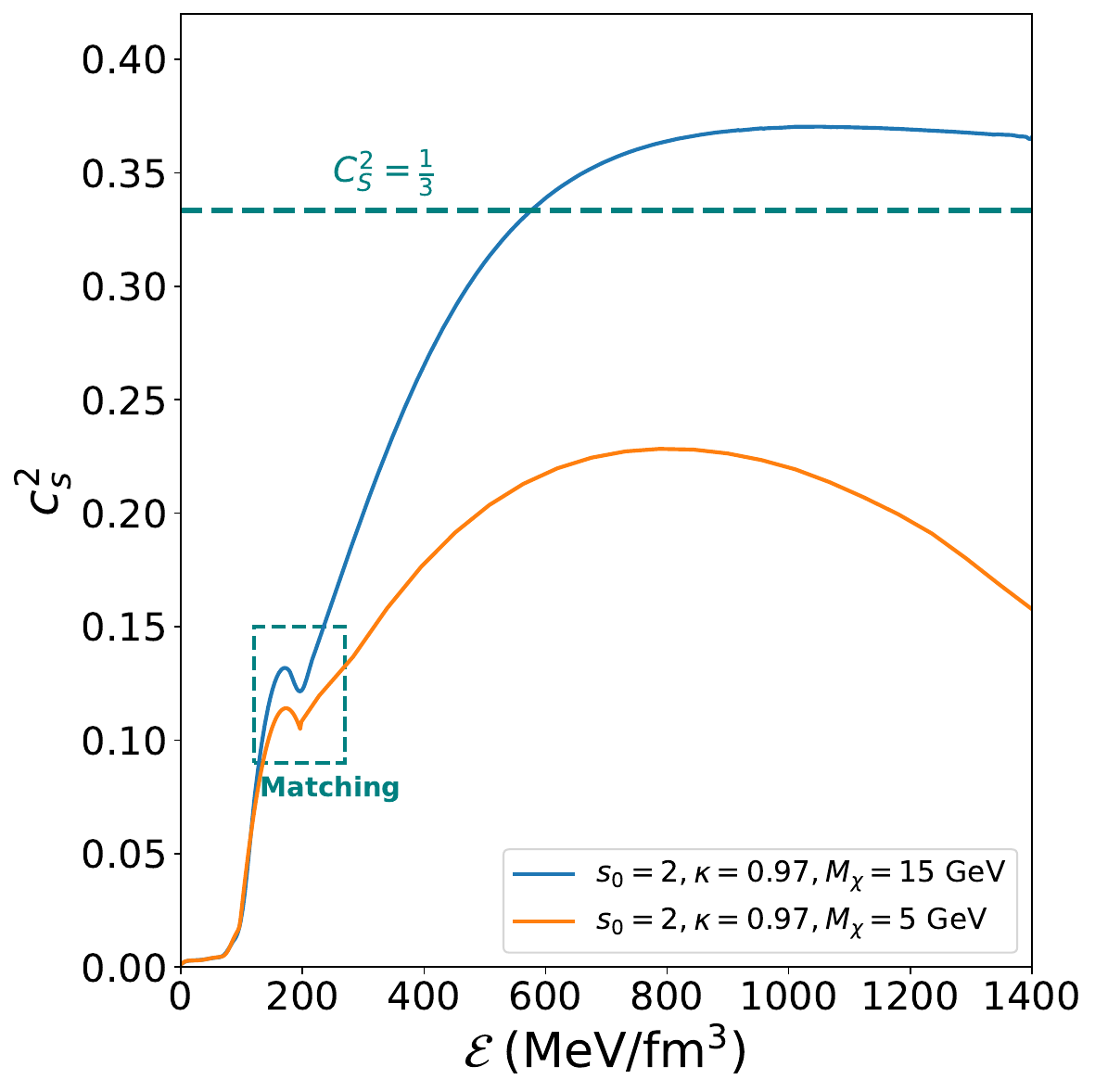}
        \par\smallskip
        (b)
    \end{minipage}

    \caption{\justifying
    Speed of sound squared ($C^2_S$) profile as function of energy density $\mathcal{E}$. (a) Dampening of the $C^2_S$ profile as $s_0$ is increased from $s_0 = 0$ to $s_0 = 2$. Inclusion of DM makes the profiles dampened even further. Moreover, it is shown that inclusion of DM makes the profile non-monotonic. (b) Dampening of $C^2_S$ profile for lighter DM particles.}
    \label{fig:E_C_2x2}
    \makeatletter
    \renewcommand{\@currentlabel}{\thefigure (a)}\label{fig:E_C_2x2_a}
    \renewcommand{\@currentlabel}{\thefigure (b)}\label{fig:E_C_2x2_b}
    \renewcommand{\@currentlabel}{\thefigure (c)}\label{fig:E_C_2x2_c}
    \makeatother
\end{figure*}

\begin{figure*}[hbt]
\centering

\begin{minipage}{0.32\textwidth}
\centering
\includegraphics[width=\linewidth]{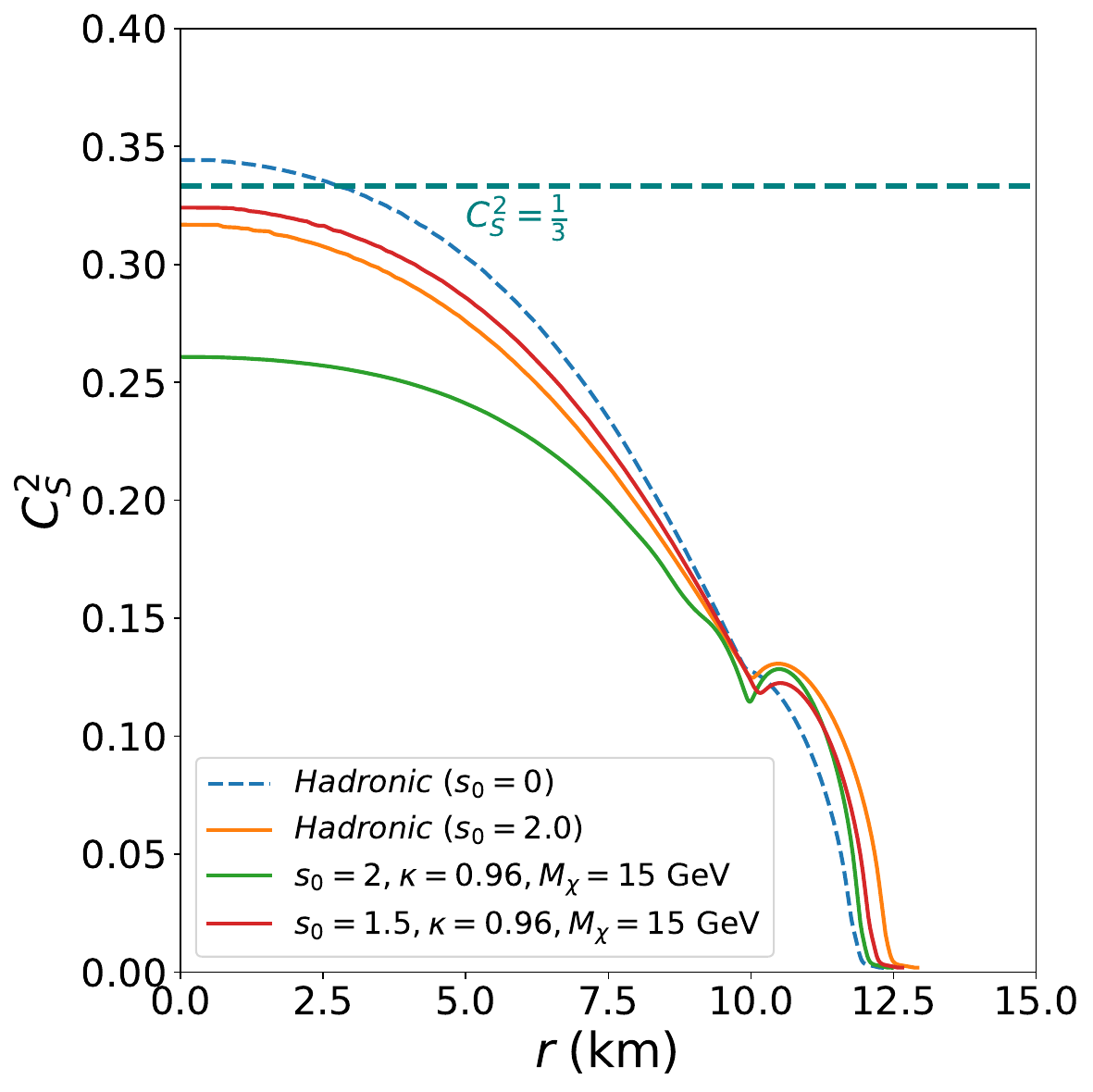}
\par\smallskip
{\small (a)}
\end{minipage}
\hfill
\begin{minipage}{0.32\textwidth}
\centering
\includegraphics[width=\linewidth]{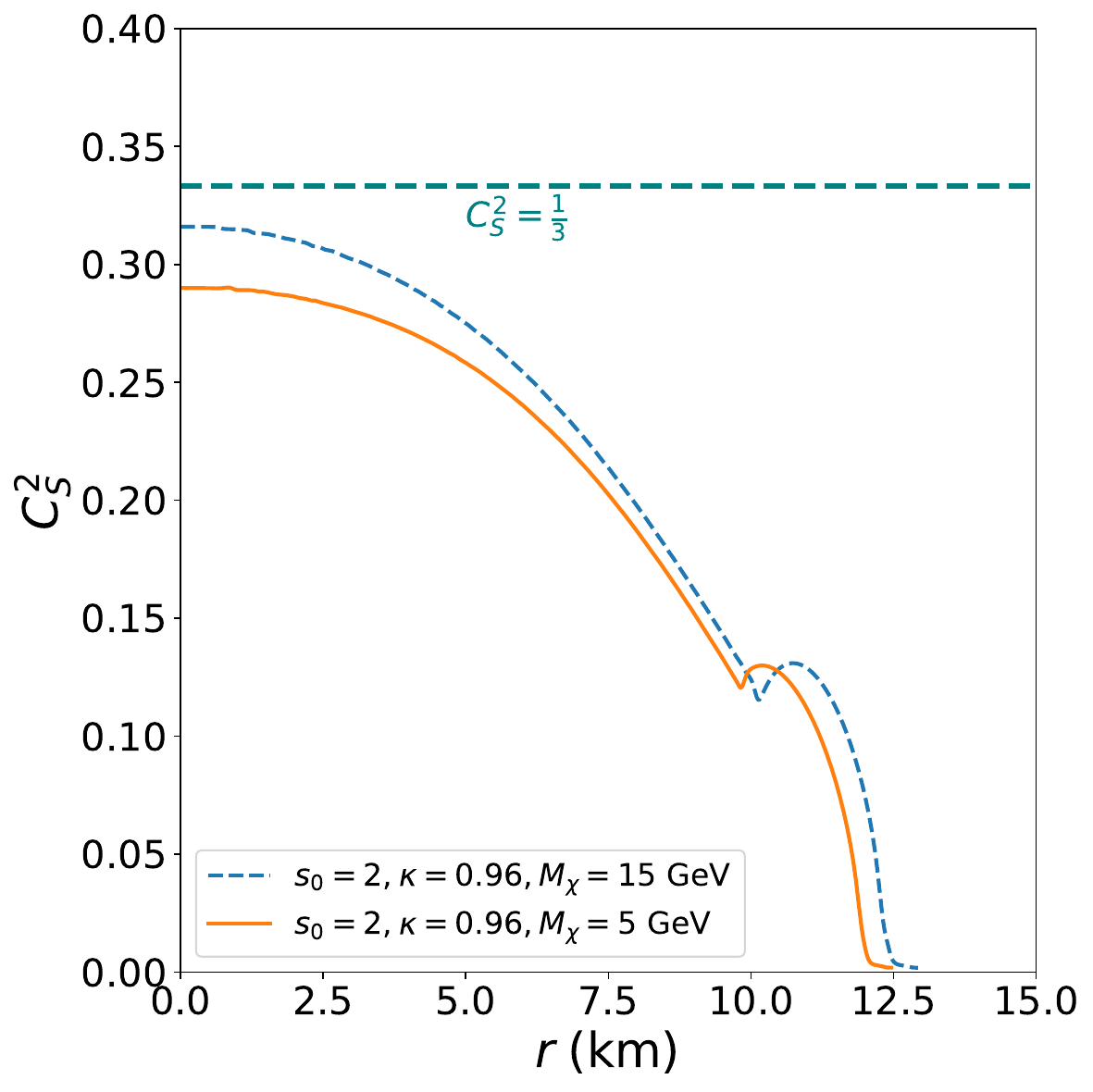}
\vspace{-0.2cm}
{\small (b)}
\end{minipage}
\hfill
\begin{minipage}{0.32\textwidth}
\centering
\includegraphics[width=\linewidth]{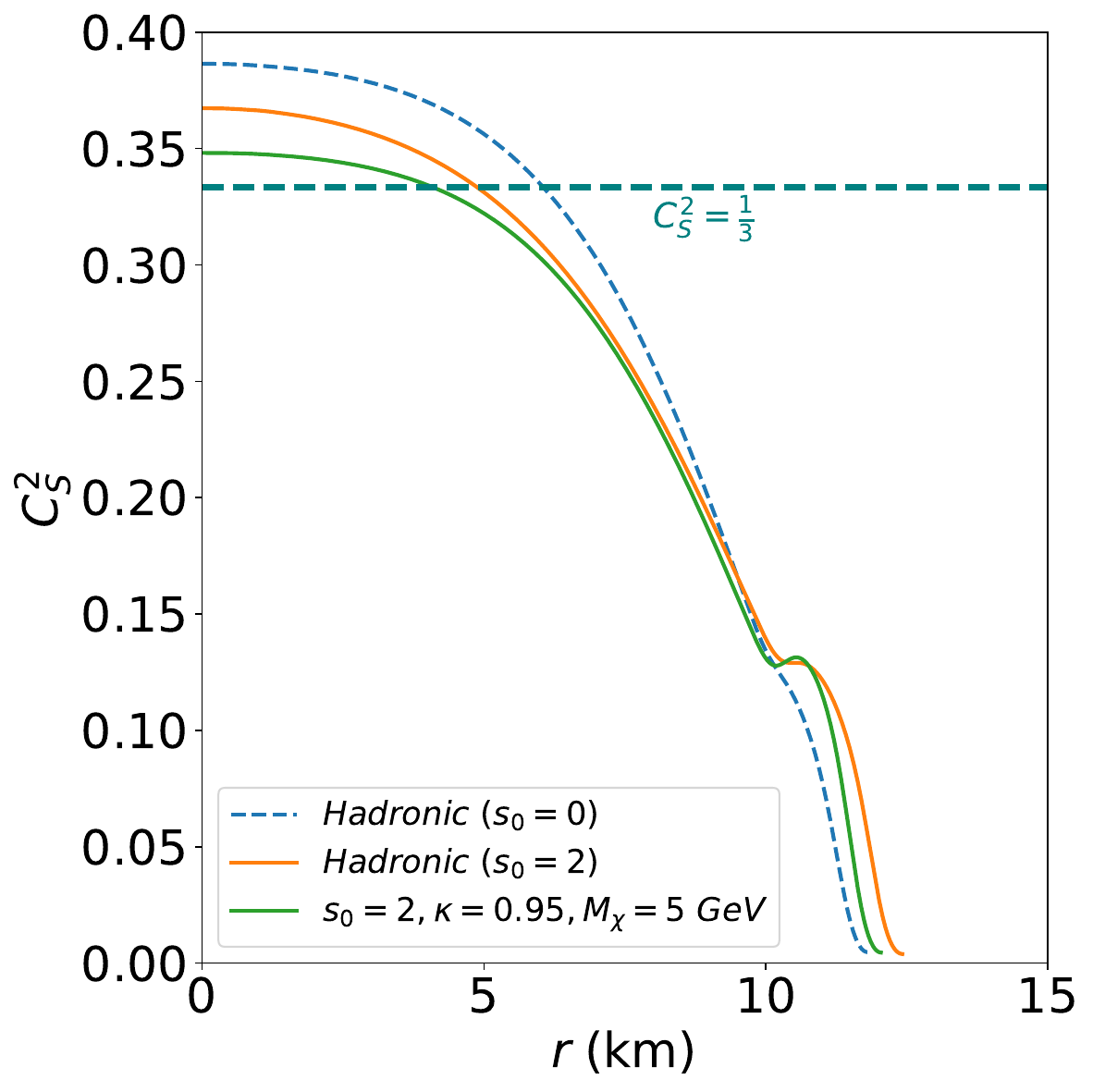}
\vspace{-0.2cm}
{\small (c)}
\end{minipage}

\caption{\justifying
Radial variation of the squared speed of sound $C_s^2(r)$ inside NS for 
(a) $M = 1.7M_\odot$, comparing hadronic cases ($s_0 = 0$ and $s_0 = 2$) with 
DM admixed cases for $M_\chi = 15~$GeV and $\kappa = 0.96$, (b) $M = 1.7M_\odot$, comparing 
cases for varying DM masses for $s_0 = 2$ and $\kappa = 0.96$, (c) $M = 1.9M_\odot$, comparing hadronic cases ($s_0 = 0$ and $s_0 = 2$) with the DM admixed case with $s_0 = 2$, $M_\chi = 5$ GeV and $\kappa = 0.95$. Comparing the results for $M = 1.7M_\odot$ 
and $M = 1.9M_\odot$, it is observed that for more massive stars, the 
conformality threshold of $C_s^2 = 1/3$ is breached near the core of 
NS for both hadronic and DM admixed cases.}
\label{fig:r_vs_cs2_combined}
\makeatletter
    \renewcommand{\@currentlabel}{\thefigure (a)}\label{fig:r_vs_cs2_combined_a}
    \renewcommand{\@currentlabel}{\thefigure (b)}\label{fig:r_vs_cs2_combined_b}
    \renewcommand{\@currentlabel}{\thefigure (c)}\label{fig:r_vs_cs2_combined_c}
    \makeatother
\end{figure*}

\begin{figure}[hbt!]
    \centering
    \includegraphics[width=\linewidth]{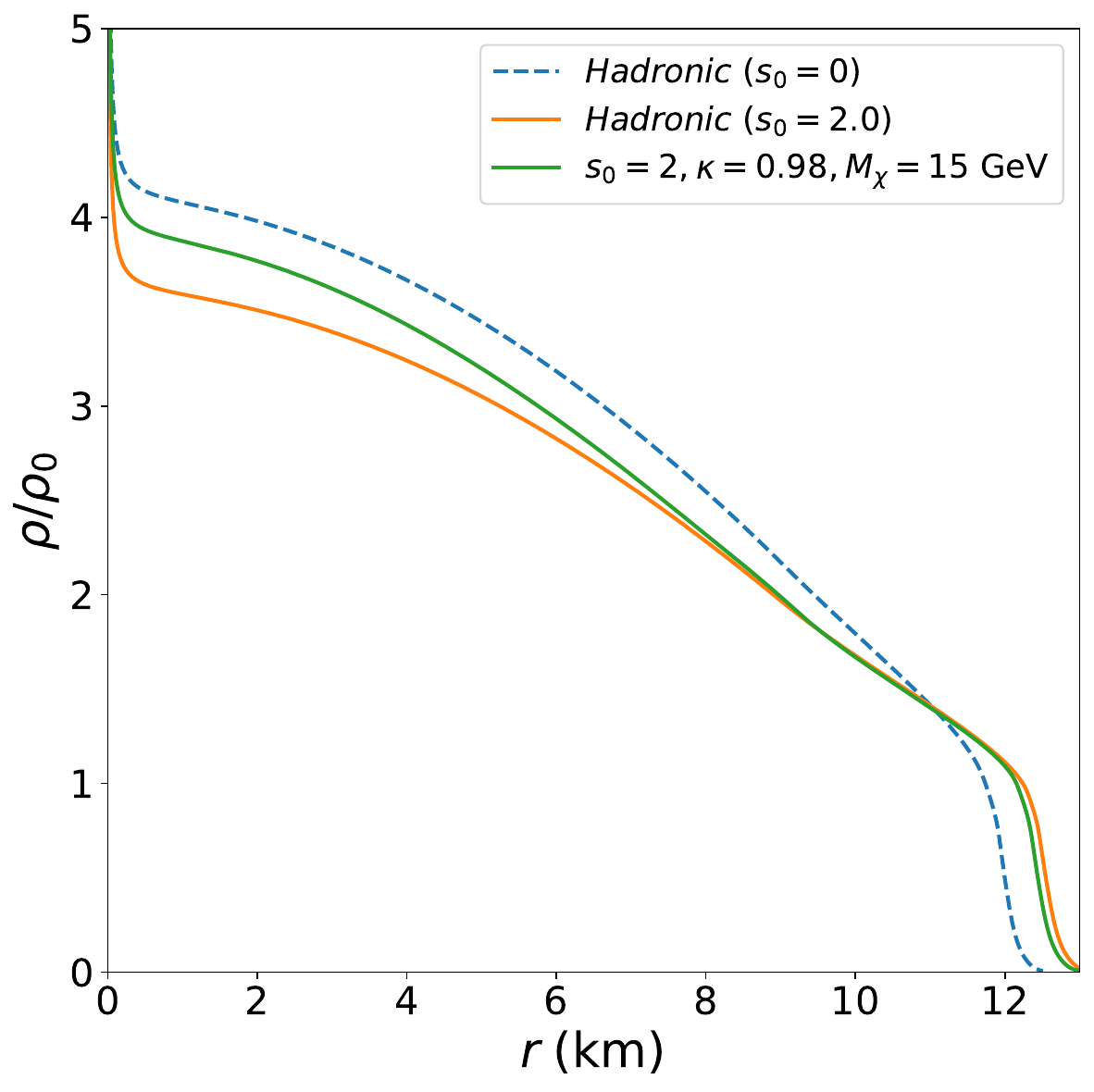}
    \caption{Radial variation of gravitational matter density (in units of nuclear saturation density) as a function of radial distance from the NS core. The competition between the thermal effects and the dark sector is apparent here. Whereas the central density decreases when $s_0$ increases from $s_0 = 0$ to $s_0 = 2$, inclusion of DM again increases the central density, similar to results obtained for cold NS in Ref. \cite{Avila:2023rzj}.}
    \label{fig:r_vs_rho}
\end{figure}

\begin{figure*}[hbt]
\centering
\begin{minipage}{0.49\linewidth}
\centering
\includegraphics[width=\linewidth]{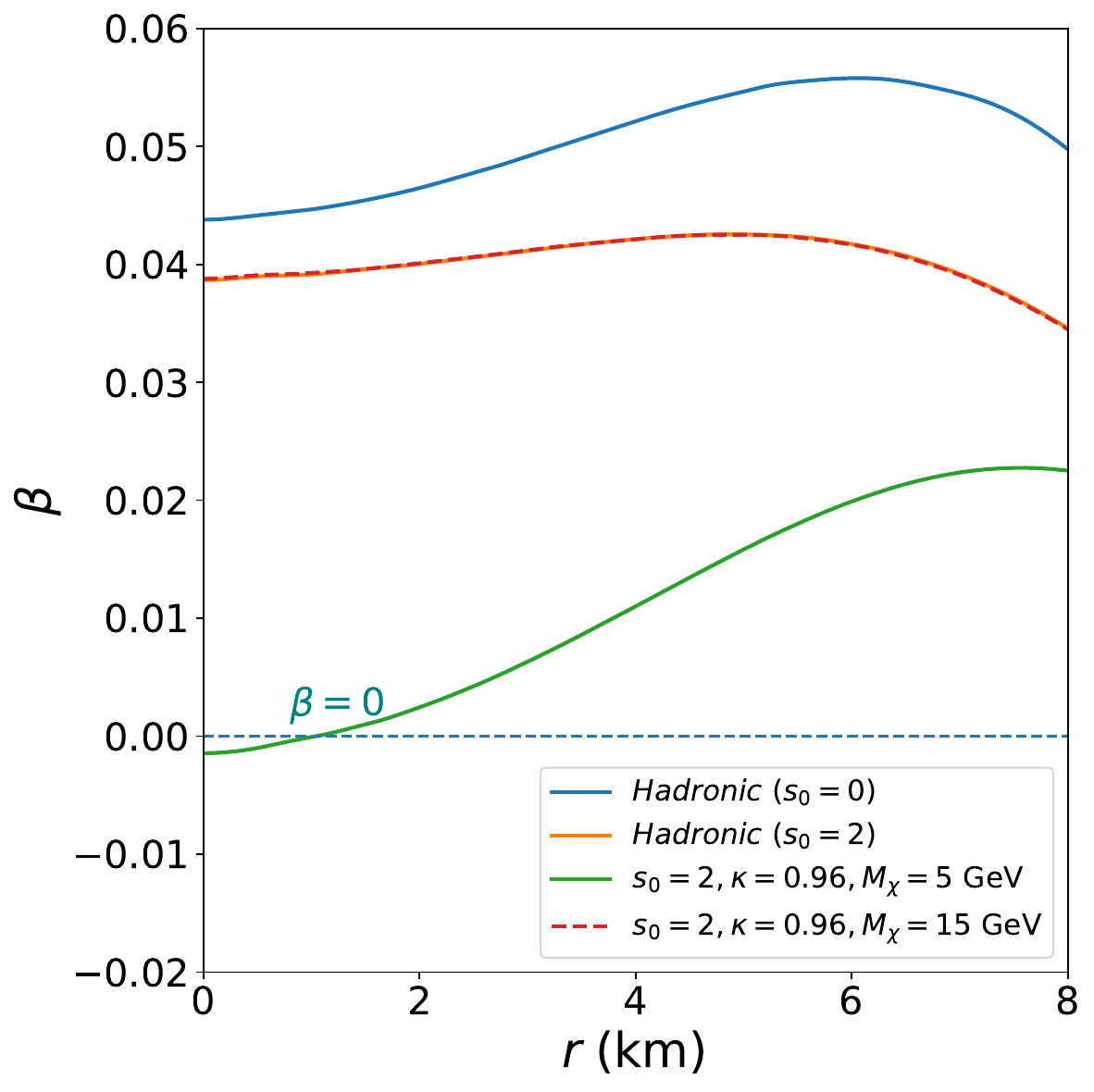}\\
(a)
\end{minipage}
\hfill
\begin{minipage}{0.49\linewidth}
\centering
\includegraphics[width=\linewidth]{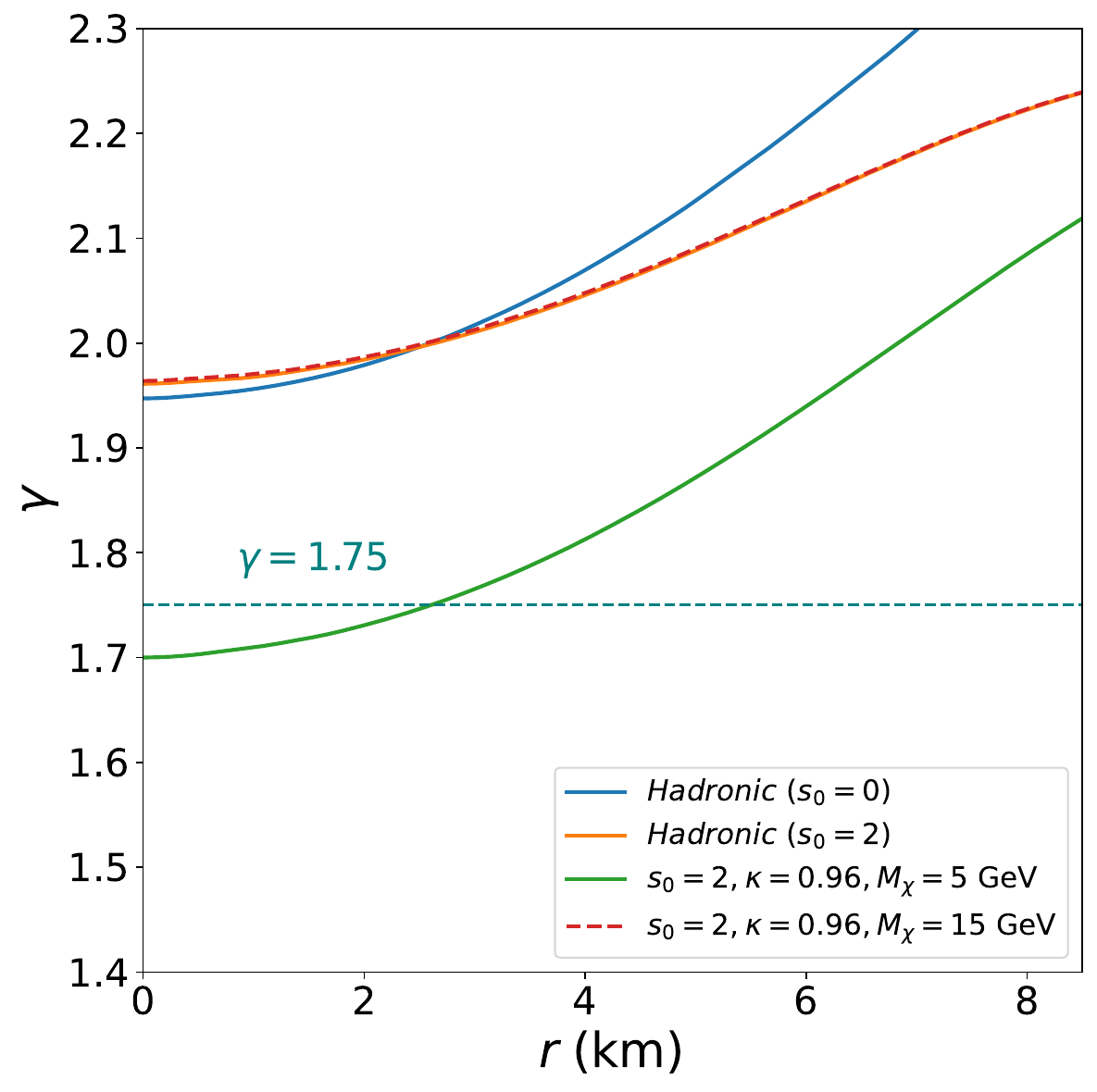}\\
(b)
\end{minipage}
\caption{\justifying
(a) Radial variation of $\beta(r)$ for $M = 1.7M_\odot$, comparing hadronic cases ($s_0 = 0, 2$) with DM admixed cases for varying $M_\chi$. Increasing $s_0$ pushes
$\beta$ closer to the conformal threshold
$\beta = 0$ compared to the purely hadronic
case ($s_0 = 0$). The DM--admixed
configuration exhibits a different behavior,
illustrating the competition between thermal
effects and DM in determining the
degree of conformality inside the star.
(b) Radial variation of the polytropic index $\gamma(r)$ for the same benchmarks. Compared to the purely hadronic
case ($s_0 = 0$), increasing $s_0$ shifts $\gamma$
closer to the conformal threshold $\gamma = 1.75$.
The inclusion of DM modifies the radial
profile of $\gamma$, illustrating the interplay
between thermal effects and DM in determining
the degree of conformality inside the star.
}
\label{fig:r_vs_betagamma_5gev}
\makeatletter
    \renewcommand{\@currentlabel}{\thefigure (a)}\label{fig:r_vs_betagamma_5gev_a}
    \renewcommand{\@currentlabel}{\thefigure (b)}\label{fig:r_vs_betagamma_5gev_b}
    \makeatother
\end{figure*}

\subsection{Parameter sets}
\label{PSET}

\noindent
The five free parameters of the hadronic sector, i.e. $C_{\sigma} = g^2_{\sigma}/m^2_{\sigma}$, $C_{\omega} = g^2_{\omega}/m^2_{\omega}$, $C_{\rho} = g^2_{\rho}/m^2_{\rho}$, $B$, and $C$, are determined by fitting to symmetric nuclear matter (SNM) properties listed in Table~\ref{tab:had_par}~\cite{Jha:2008yth}. The tabulated values of SNM properties e.g. saturation density ($\rho_0 = 0.153$~fm$^{-3}$), nucleon effective mass at saturation [$Y(\rho_0) = 0.87$], nuclear incompressibility ($K\approx 231$ MeV), binding energy per nucleon ($E_{BE}/A = -16.3$~MeV) and symmetry energy coefficient ($J = 32$~MeV), calculated using the specified parameter set, agree well with already existing constraints. The value of the slope parameter at saturation ($L_0=88$) also falls well within already established constraints~\cite{Dutra:2014qga}. 

We use the parameter set for the DM sector presented in Table~\ref{tab:dark_par}. These values fall very well in agreement with the self-interaction cross-section constraints from the Bullet Cluster and the present thermal relic abundance of DM~\cite{Tulin:2013teo}. The couplings of the new mediators with the nucleons, $g_{\phi}$ and $g_{\xi}$, are much smaller than the corresponding couplings with the DM, $y_{\phi}$ and $y_{\xi}$. Therefore, the DM-nucleon scattering cross section agrees well with the available constraints~\cite{Randall:2008ppe}. In this work, we only consider NS composed of $npe\chi$ matter with the $npe$ sector subjected to charge neutrality and $\beta$-equilibrium conditions. Moreover, we will use $s_0 = 1.5$ and 2 as the benchmark values for entropy per baryon for different values of the parameter $\kappa = \mu_\chi/M_\chi$.

The choice of $s_0 = 1.5,~2$ chosen in this work reflects the characteristic range of $s_0$ found in detailed transport simulations of newly formed and evolving NSs. Core-collapse simulations \cite{Burrows:1986me,Pons:1998mm} show that a proto-neutron star is born with $s_0\approx~1$ and increases to $s_0 \approx~2$ over the subsequent Kelvin-Helmholtz deleptonization phase. This range is therefore the well-motivated window for hot compact object studies \cite{Prakash:1996xs,Issifu:2025gsq,Issifu:2025jac,Issifu:2024htq,Issifu:2025qqw}.
The values of $s_0$ ($= 1.5, 2$) and $\kappa$ have been chosen such that the resulting deviations from the purely hadronic EoS are sufficiently pronounced to be observable. We find that for smaller values of $s_0$, GeV-scale dark matter does not induce any appreciable modifications to the EoS and, consequently, leads to negligible observable effects. 

\subsection{Temperature Profiles in the Presence of Dark Matter}
\label{Temp_Profile}

\noindent
In the relativistic mean field (RMF) approximation, one replaces the meson fields by their RMF expectation values and, hence, all the corresponding kinetic terms drop out of the RMF approximated model. The effective masses of nucleons and the DM are then given by $m_N^{\ast} = g_{\sigma}\sigma + g_{\phi}\phi$ and $m^{*}_{\chi} = M_{\chi} + y_{\phi}\phi$. However, as the RMF EoM need to be solved together with the constraint of constant entropy per baryon $s$ ($= s_0$ being a fixed value of the constant entropy per baryon) so that the temperature profile as a function of the total baryon density $\rho^V_{\text{tot}}$ can be obtained, the solutions of the mean field EoM are functions of both $\rho^V_{\text{tot}}$ and $s_0$. A detailed discussion on notations, EoM, EoS and their representations in terms of generalized Fermi-Dirac integrals (GFDI) is presented in Appendices \ref{app:eos_eom} and \ref{GFD_Observables}. In Fig. \ref{fig:T_vs_Rho_a}, we show temperature profiles $T(\rho^{V}_{\text{tot}})$ as functions of baryon density $\rho^V_{\text{tot}}$ for fixed $s_0 = 2$, $\kappa = 0.97$ and for $M_\chi = 5~\text{GeV}$ and $M_\chi = 15~\text{GeV}$. As $s_0 = 0$ corresponds to the profile $T(\rho^V_{\text{tot}}) = 0$, its explicit profile has not been shown in the figure. It is evident that for non-zero $s_0$, the inclusion of DM cools the NS core, supported by \cite{Bhat:2019tnz} as opposed to the results shown in \cite{Issifu:2025qqw,Issifu:2025jac} and in support of the result in \cite{Bell:2023ysh}. One can see a competition between thermal effects due to non-zero $s_0$ and the cooling effects due to the presence of DM. Compared to the hadronic case with $s_0 = 0$, whereas increasing $s_0$ amplifies the temperature profile, the addition of DM further decreases the profile. Moreover, for the same $\kappa$, lighter DM particles contribute more to the cooling. In Fig. \ref{fig:T_vs_Rho_b}, we show $T(\rho^{V}_{\text{tot}})$ for $s_0 = 2$ and $M_\chi = 15~\text{GeV}$ for varying $\kappa$. An increase in $\mu_\chi$ leads to a cooler NS core and an overall decrease in stellar temperature.

\section{Equation of state}
\label{Equation_of_State}

\noindent
The expressions for the total pressure and energy density of DMANS matter are given as follows:
\begin{align}
\label{eq:PE1}
\mathcal{P}
&= \sum_{f = N, \chi, \ell} \mathcal{P}_f
 + \sum_{f = \overline{N}, \overline{\chi}, \overline{\ell}} \overline{\mathcal{P}}_f
 - \Sigma + \Theta_{-} ,
 \end{align}
\begin{align}
\label{eq:PE2}
\mathcal{E}
&= \sum_{f = N, \chi, \ell} \mathcal{E}_f + \sum_{f = \overline{N}, \overline{\chi}, \overline{\ell}} \overline{\mathcal{E}}_f
 + \Sigma + \Theta_{+} .
\end{align}
The expressions for the quantities $\Sigma$, $\Theta_{\pm}$, $\mathcal{P}_f$, $\overline{\mathcal{P}}_f$, $\mathcal{E}_f$ and $\overline{\mathcal{E}}_f$ can be found in Appendices \ref{app:eos_eom} and \ref{GFD_Observables}. In the above expressions, the solutions of the RMF fields and the temperature profile $T = T(s_0, \rho^V_{\text{tot}})$ should be obtained from self-consistent solutions of Eqs.~\eqref{eom:phi} - \eqref{eom:sigma} together with the condition of charge neutrality, chemical equilibrium and constancy of entropy per baryon for fixed $\kappa$. These solutions, in turn, are used to numerically evaluate the DM density $\rho^{V}_{\chi}(s_0, \rho^{V}_{\text{tot}}$) as a function of $s_0$ and $\rho^V_{\text{tot}}$. 

\par As the temperature of the crust of the NS is expected to be much smaller compared to the core, we use the following procedure of matching BSK21 EoS at the crust: at the saturation density $\rho = \rho_0 = 0.15$~fm$^{-3}$, we choose a range of energy density $(\mathcal{E}(\rho_0) - \Delta \mathcal{E}, \mathcal{E}(\rho_0) + \Delta \mathcal{E})$ (in this work, we have chosen $\Delta\mathcal{E} \approx 0.3~\mathcal{E}(\rho_0)$). For $\mathcal{E} < \mathcal{E}(\rho_0) - \Delta \mathcal{E}$, we use the BSK21 as the crust of the EoS; for $\mathcal{E} > \mathcal{E}(\rho_0) + \Delta \mathcal{E}$, we use the warm DMANS EoS obtained from the model and within the range $[\mathcal{E}(\rho_0) - \Delta \mathcal{E},\mathcal{E}(\rho_0) + \Delta \mathcal{E})]$; we impose continuity conditions for $\mathcal{P}$ and the speed of sound squared $C^2_S$ at $\mathcal{E} = \mathcal{E}(\rho_0) - \Delta\mathcal{E}$ and $\mathcal{E} = \mathcal{E}(\rho_0) + \Delta\mathcal{E}$. These four conditions have been used to uniquely specify the EoS inside the range $[\mathcal{E}(\rho_0) - \Delta \mathcal{E}, \mathcal{E}(\rho_0) + \Delta \mathcal{E})]$ as a simple phenomenological cubic EoS of the form $\mathcal{P}(\mathcal{E}) = a + b\,\mathcal{E} + c\,\mathcal{E}^2 + d\,\mathcal{E}^3$.

\par The DM density profiles as functions of the total baryon density $\rho^{V}_{\text{tot}}$ for parameter sets BPI and BPII are shown in Fig.~\ref{fig:dm_eos_combined_a}. There are three quantities, $s_0$, $M_\chi$ and $\kappa$, that affect the DM density profiles. For fixed $M_\chi$ and $\kappa$, a higher value of $s_0$ amplifies the DM density profile, which contributes to the softening of the EoS, as seen in Fig.~\ref{fig:dm_eos_combined_b}. As an example, for $M_\chi = 15~$GeV, the profile for $s_0 = 2$ is more heightened compared to the profile for $s_0 = 1.5$ for both $\kappa = 0.97$ and $\kappa = 0.98$. This is reflected in Fig.~\ref{fig:dm_eos_combined_b}: the EoS for $s_0 = 2$ is much softer for the same values of $\kappa$ and $M_\chi$ compared to the EoS for $s_0 = 1.5$.

\par For fixed $s_0$ and $\kappa$, lighter DM particles have enhanced density profiles compared to heavier ones, which again contributes to softening the EoS. For instance, it is clear that for $s_0 = 2$ and $\kappa = 0.97$, the DM profile for $M_\chi = 5~$GeV is much more heightened compared to the profile for $M_\chi = 15~$GeV. This is reflected in Fig. \ref{fig:dm_eos_combined_b}, where for the same values of the parameters, the EoS for $M_\chi = 5$~GeV is much softer compared to the EoS for $M_\chi = 15$~GeV. For $s_0 = 2$, $M_\chi = 5$~GeV and $\kappa = 0.97$, the slope of the EoS is visibly non-monotonic. This non-monotonicity will carry over to the speed of sound profile, akin to what is suggested for cold NS EoS \cite{Kojo:2020krb,Ecker:2017fyh,Altiparmak:2022bke,Brandes:2022nxa,Roy:2022nwy}. 

Lastly, for fixed $s_0$ and $M_\chi$, an increasing value of $\kappa$ amplifies the DM density profiles. This becomes evident by comparing results for $s_0 = 2$, $M_\chi = 15$~GeV, $\kappa = 0.97$ and $s_0 = 2$, $M_\chi = 15$~GeV, $\kappa = 0.98$ in Fig.~\ref{fig:dm_eos_combined_a}. This is also reflected in Fig.~\ref{fig:dm_eos_combined_b} where the EoS for the latter is much softer compared to the former. The DM profiles presented in Fig.~\ref{fig:dm_eos_combined_a} suggest that for increasing $\rho^{V}_{\text{tot}}$ (and hence for increasing value of $\mathcal{E}$), $\rho_{\chi}$ also increases, and hence the degree by which the EoS softens will increase as $\mathcal{E}$ increases. This will become evident in the non-monotonic profile for the speed of sound as a function of $\mathcal{E}$. 

\section{Macroscopic Observables}
\label{Result}

\par In this section, we present predictions for macroscopic observables. Fig.~\ref{fig:RxMy_2x4} shows the $M$-$R$ relations. Predictions are also compared with observations of various compact objects and $\Lambda(1.4M_\odot) < 400$ constraint on the $M$-$R$ plane as obtained from Ref.~\cite{Annala:2017llu}. The qualitative nature of $M$-$R$ relations is similar to the results found in Ref.~\cite{Issifu:2025jac}, which uses non-annihilating, asymmetric fermionic and bosonic DM that interacts with the hadronic sector only via gravity.
\par It is evident from the figures that for a fixed $s_0$, the low-mass part of the $M$-$R$ relation is practically unchanged and coincident. This is because at relatively low energy density, the DM density is too small to have any noticeable effect on the EoS, and hence, on the $M$-$R$ relations. However, in the range $M = 1.25 - 2.0M_\odot$, a substantial effect due to enhanced DM density profiles manifests itself as a reduction of the maximum NS mass. We have observed that as $\kappa$ increases, the maximum mass decreases due to softening of the EoS. For instance, in Fig.~\ref{fig:RxMy_2x4_a}, whereas for the purely hadronic case with $s_0 = 2$ the maximum mass of the NS is around $1.98M_\odot$, for $\kappa = 0.98$, and $M_\chi = 15$~GeV, the maximum mass drops to $1.77M_\odot$. Table~\ref{tab:BP1_BP2_mixedalpha} shows, for $s_0 = 2$, maximum stellar mass $M_\text{max}$, corresponding stellar radius $R_\text{max}$ and tidal polarizability at $M = 1.4M_\odot$, $\Lambda_{1.4}$ for both BPI and BPII sets.
\par Increasing $s_0$ is seen to increase the maximum mass of the NS. For example, in Fig.~\ref{fig:RxMy_2x4_a} for the purely hadronic cases when $s_0$ is increased from $s_0 = 0$ to $s_0 = 2$. On the other hand, this mass is pulled down to $M_{\text{max}}\approx 1.77M_\odot$ upon inclusion of DM with $M_\chi = 15~$GeV and $\kappa = 0.98$. This is because, as we have seen before for the EoS, a finite value of $s_0$ makes the EoSs soften due to DM accumulation inside NS, which reduces $M_{\text{max}}$. This essentially indicates a competition in which thermal effects and the softening effect of the DM pull the maximum masses in opposite directions, as seen in the temperature profiles in Fig.~\ref{fig:T_vs_Rho_Combined}. In the regime of sufficiently high masses, at a particular mass of the star, as $\kappa$ increases and more DM is accumulated in the core, the radius of the star decreases and the compactness of the star increases as a function of stellar mass.

\par In Fig.~\ref{fig:M_Lambda_2x2}, we present the mass-tidal deformability ($M$-$\Lambda$) relations for the same parameter sets. The competition between the thermal effects and the dark sector is apparent. In Fig.~\ref{fig:M_Lambda_2x2_a}, whereas $\Lambda(1.4M_\odot)$ increases when $s_0$ is increased from $s_0 = 0$ ($\Lambda_{\text{1.4}}\approx 640$) to $s_0 = 2$ ($\Lambda_{\text{1.4}}\approx 880$), inclusion of DM ($M_\chi = 15~$GeV and $\kappa = 0.98$) again reduces its value ($\Lambda_{\text{1.4}}\approx 814$), again highlighting the competition between the two sectors. For a fixed $s_0$ and $\kappa$, higher DM mass leads to a higher value of $\Lambda$ for a fixed stellar mass, as shown in Fig.~\ref{fig:M_Lambda_2x2_b}. As an example, for $s_0 = 2$ and $\kappa = 0.97$, where $\Lambda_{1.4}\approx 881$ for $M_\chi = 15~$GeV, the same value reduces to $\Lambda_{1.4}\approx 511$ for $M_\chi = 5~$GeV. 

\par Fig.~\ref{fig:MI_RI_combined} represents mass-moment of inertia ($M$-$I$) relations. One can identify the same kind of competition between thermal effects and the DM sector as well. For sufficiently high stellar masses, whereas the effect of increasing $s_0$ from $s_0 = 0$ to $s_0 = 2$ is to increase the MoI, inclusion of DM again reduces it, most notably above $M = 1.4M_\odot$, as can be seen from Fig.\ref{fig:MI_RI_combined_a}. The same trend is true for the maximum value of MoI. Also, for fixed $s_0$ and $\kappa$, larger $M_\chi$ leads to larger value of MoI for all stellar masses, as shown in Fig.~\ref{fig:MI_RI_combined_b}.

\section{Speed of Sound}
\label{SOS_Profile}

Fig.~\ref{fig:E_C_2x2} presents $C^2_S-\mathcal{E}$ relations. The speed of sound ($C^2_S$) inside an NS is a direct measure of the stiffness of the EoS. This is defined as the slope of the EoS on an isentrope via
\begin{align}
    C^2_S = \bigg(\frac{d \mathcal{P}}{d \mathcal{E}}\bigg)_s
\end{align}
Although $C^2_S$ cannot be directly measured, it can be inferred indirectly from EoS constraints and other NS observables obtained through pulsar and GW data. In contrast to the case for cold NS where constraints on $C^2_S$ are available \cite{Mroczek:2023zxo,Yao:2023yda,Chatterjee:2023ecc,Brandes:2022nxa,Marczenko:2024uit,Altiparmak:2022bke,Kojo:2020krb,Yao:2023yda,Chen:2023hqm,Itou:2022ebw,Ferrer:2022afu,Aloy:2018jov,Bedaque:2014sqa}, apart from causality $0 \leq C^2_S \leq 1$ there are no constraints on $C_S^2$ for warm DMANS matter, keeping fair amount of flexibility in speed of sound profile for such warm systems. In Fig.~\ref{fig:E_C_2x2}, we show the $C^2_S$ profile as a function of energy density $\mathcal{E}$. The line $C^2_S = 1/3$ denotes the conformal limit. The $C^2_S$ profile is dampened both by increasing $s_0$ and by adding DM to the system. However, in the absence of DM in the system, as $s_0$ is changed from $s_0 = 0$ to $s_0 = 2$, the profile of $C^2_S$ does not develop any non-monotonicity. For fixed $M_\chi$ and $\kappa$, a higher value of entropy per baryon $s_0$ leads to lower values of $C^2_S$, as can be seen from Fig.~\ref{fig:E_C_2x2_a}. On the other hand, for fixed $s_0$ and $\kappa$, higher $M_\chi$ leads to a higher value of $C_S^2$ for all values of $\mathcal{E}$ greater than the matching point value, as can be seen from Fig.~\ref{fig:E_C_2x2_b}. Increasing $M_\chi$ assists the system to breach the conformal limit, whereas increasing $s_0$ inhibits it as it dampens the $C^2_S$ profile. So whether the system will breach the conformal limit depends on this competition mechanism between thermal effects and the dark sector. The non-monotonic profile of $C^2_S$ \cite{Das:2025skn} becomes apparent and visible as $\kappa\rightarrow 1$
, similar to inferred in numerous works for cold NS matter \cite{Mroczek:2023zxo,Yao:2023yda,Brandes:2022nxa,Marczenko:2024uit,Altiparmak:2022bke,Kojo:2020krb,Yao:2023yda,Chen:2023hqm,Itou:2022ebw,Ferrer:2022afu,Aloy:2018jov,Bedaque:2014sqa} and contrary to the results in Ref.~\cite{Issifu:2025qqw}. However, in such cases, as increasing $\kappa$ decreases the maximum value of $C^2_S$, and, as a result, $C^2_S$ may not be able to breach the conformal limit. This is apparent from both Fig.~\ref{fig:E_C_2x2_a} and Fig.~\ref{fig:E_C_2x2_b}. 

\section{Radial Variations and Conformality Indicators}
\label{radial_variations}

\par In Fig.~\ref{fig:r_vs_cs2_combined}, we present the radial variation of the speed of sound $C^2_S(r)$: for $M = 1.7M_\odot$ NS in Figs.~\ref{fig:r_vs_cs2_combined_a} and \ref{fig:r_vs_cs2_combined_b} and for $M = 1.9M_\odot$ in Fig.~\ref{fig:r_vs_cs2_combined_c}. Comparing Figs.~\ref{fig:r_vs_cs2_combined_a} and \ref{fig:r_vs_cs2_combined_c}, we observe that heavier stars can sustain greater values of $C^2_S(0)$ at the NS core. Moreover, for the same $M_\chi$, $\kappa$ and the stellar mass, a higher value of $s_0$ leads to a lower value of $C^2_S(0)$, as can be clearly seen by comparing Fig.~\ref{fig:r_vs_cs2_combined_a} and Fig.~\ref{fig:r_vs_cs2_combined_b} and also corroborated by the previous results in Fig.~\ref{fig:E_C_2x2}. For fixed $s_0$, $\kappa$ and stellar mass, larger value of $M_\chi$ particles lead to higher $C^2_S(0)$, as can be observed from Fig.~\ref{fig:r_vs_cs2_combined_b}. However, none of the profiles among the purely hadronic cases for non-zero $s_0$ and DM admixed cases in Figs.~\ref{fig:r_vs_cs2_combined_a} and \ref{fig:r_vs_cs2_combined_b} with $M_\odot = 1.7M_\odot$ breach the conformal limit of $C^2_S = 1/3$ at the core of NS. Such a breach of the conformal limit for both hadronic and DM admixed cases can be achieved for sufficiently massive stars, as is shown for $M = 1.9M_\odot$ in Fig.~\ref{fig:r_vs_cs2_combined_c}. As increasing $\kappa$ decreases $C^2_S(0)$, this puts limits on $\kappa$, for a particular $M_\chi$, $s_0$ and stellar mass, above which the conformal limit is not breached. In such cases of heavy NSs, increasing $\kappa$ makes the system hit the conformal threshold closer to the NS core, as can be seen from Fig.~\ref{fig:r_vs_cs2_combined_c}. \par Fig.~\ref{fig:r_vs_rho} shows radial profile of the gravitational matter density for a $M = 1.4M_\odot$ NS. The competition between the thermal effects and the dark sector is apparent here. For instance, whereas increasing $s_0$ from $s_0 = 0$ to $s_0 = 2$ decreases matter density near the core, including DM of mass $M_\chi = 15~$GeV for $\kappa = 0.98$ increases it again, similar to the results reported in Ref.~\cite{Avila:2023rzj} for cold NSs. A similar trend has been observed for radial variation of energy density, as previously obtained in the context of asymmetric DM \cite{Gresham:2018rqo}.
\par In Fig.~\ref{fig:r_vs_betagamma_5gev_a}, we show the radial variation of curvature of energy per particle $\beta(r)$ as a function of radial distance from the NS core for $M = 1.7M_\odot$.  Increasing the value of $s_0$ leads the value of $\beta$ closer to the conformality threshold, as can be seen from comparing purely hadronic cases in the figure. The same effect has been observed for fixed $M_\chi$, $\kappa$ and stellar mass. On the other hand, for fixed $s_0$, $\kappa$ and stellar mass, heavier DM pushes the system away from the conformal threshold. Lighter DM particles actually assist the system in breaching the conformal threshold for warm systems. For instance, whereas for $s_0 = 2$, $\kappa = 0.96$ and $M_\chi = 15$~GeV, the system does not breach the conformal threshold for $\beta$, for $M_\chi = 5$~GeV, conformal threshold is breached at $r = 1.12~\text{km}$ away from the stellar core. A similar trend for the polytropic index $\gamma(r)$ can be seen in 
Fig.~\ref{fig:r_vs_betagamma_5gev_b}. For $s_0 = 2$, $\kappa = 0.96$, whereas for $M_\chi = 15$~GeV, the system does not breach the conformal limit, for $M_\chi = 5$~GeV, the conformal threshold $\gamma \approx 1.75$ at $r \approx 2.6$~km away from the stellar core. Similar trends have been observed for the conformal distance $d_C$.

\section{Conclusion and Discussion}
\label{END}

In this work, we investigate the effect of self-consistent density profiles of GeV-scale DM  and stellar temperature profiles on macroscopic observables and conformal behavior of isentropic DMANS with a hot core and relatively cold crust. We use the RMF effective chiral model to describe the $npe$ matter under charge neutrality and $\beta$-equilibrium conditions. For the dark sector, we use GeV-scale fermionic DM with a scalar and a vector mediator connecting the two sectors. In order to take into account self-consistent temperature and DM density profiles, EoM are self-consistently solved under fixed entropy per baryon.

\par We observe that in the presence of DM in the system, increasing the value of the entropy per baryon amplifies the DM density profile, which contributes to the gradual softening of the EoS as energy density increases. On the other hand, lighter DM particles lead to enhanced density profiles compared to heavier ones for the same entropy per baryon, which again contributes to softening the EoS. At low energy densities, the DM density is too small to have any significant effect on the $M$-$R$ relations. However, for sufficiently massive stellar configurations, noticeable effects due to the sufficient DM accumulation manifest themselves as a reduction of the maximum mass. A competition is observed between the thermal effects and the softening effects of the dark sector. Whereas the maximum mass increases by increasing the value of entropy per baryon, further inclusion of DM reduces it. This is because for finite entropy per baryon, inclusion of DM softens the EoS for sufficiently high energy densities, which reduces the maximum mass of NS. On the other hand, larger DM masses lead to stiffer EoS, and hence, lead to larger maximum mass of the star. For all the macroscopic observables, this competition between the thermal and the dark sector is consistently observed. Given this interplay, it is plausible that certain combinations of entropy per baryon and DM mass may yield stellar properties that closely resemble those of the cold hadronic scenario for maximum mass and $\Lambda_{\text{1.4}}$.

\par The speed of sound ($C^2_S$) profile as a function of energy density shows that, whereas higher DM mass leads to a larger overall value of $C^2_S$, a higher value of entropy per baryon leads to lower values of $C^2_S$, again indicating the competition between these two sectors. Whether the DM admixed system will breach the conformal limit actually depends on a delicate balance between the entropy per baryon, DM mass and DM chemical potential. The non-monotonic nature of $C^2_S$ for the DM admixed cases becomes much more apparent as the DM chemical potential approaches the DM mass. For the radial variation $C^2_S(r)$, curvature of energy per particle $\beta(r)$~\cite{Marczenko:2023txe} and polytropic index $\gamma(r)$~\cite{Annala:2019puf}, it is observed that for sufficiently massive stars, these measures can breach their conformal threshold values sufficiently close to the NS core for DM admixed cases. The radial profiles $\beta(r)$ and $\gamma(r)$ demonstrate similar competition between thermal and dark sectors. 

\par There is an important subtlety in the results of conformality. The approach toward conformality is driven not solely by the hadronic matter reaching extreme densities, but through a delicate balance between thermal effects and DM accumulation. In particular, lighter DM particles accelerate the approach of warm NS matter toward conformality. This implies that the conformality signatures attributed to quark-matter cores in cold NS could be mimicked by DM admixture in isentropic stars. Distinguishing between these two cases is essential and necessary for multi-messenger observations to independently constrain both the warm hadronic state and the DM content of massive compact objects.
\section*{acknowledgment}
P.J. is supported by a grant from the U.S. National Science Foundation PHY-2310003. T.M. acknowledges partial support from the SERB/ANRF,
Government of India, through the Core Research Grant No. CRG/2023/007031. 
\appendix
\section{EoS and EoM}
\label{app:eos_eom}

\noindent
The quantity $\Sigma$ capturing the contribution from the scalar potential $U(\bar{x})$, is given by:
\begin{align}
\label{Cap_sigma}
\Sigma = m^{2}_{\sigma}f^{2}_{\pi}\left[\frac{\overline{Y}^4}{8} - \frac{f^2_{\pi}\overline{Y}^6}{12}B + \frac{f^4_{\pi}\overline{Y}^8}{16}C\right],
\end{align}
where $\overline{Y}^2=1-Y^2$ where $Y = Y(s_0, \rho)$. The quantity $\Theta_{\pm}$ is
\begin{align}
\label{Cap_Theta}
\Theta_{\pm} = \frac{C_{\omega}{\lt(\rho_{\text{tot}}^V\rt)}^2}{2Y^2} + \frac{1}{2}m^2_{\rho}\lt(\rho^{03}\rt)^2 + \frac{1}{2}m^2_{\xi}\lt(\xi^{0}\rt)^2 \pm \frac{1}{2}m^2_{\phi}\phi^2,
\end{align}
The EoMs for $\phi$, $\xi^0$, $\omega^0$ and $\rho^{03}$ in the mean field approximation are given as follows
\begin{equation}
    \phi = - \frac{g_{\phi}}{m^2_{\phi}} \rho_{\text{tot}}^{S} - \frac{y_{\phi}}{m^2_{\phi}} \rho^{S}_{\chi, \text{tot}},
    \label{eom:phi}
\end{equation}

\begin{align}
    \xi^{0} &= \frac{g_{\xi}}{m^2_{\xi}}\rho_{\text{tot}}^V +\frac{y_{\xi}}{m^2_{\xi}}\rho^{V}_{\chi, \text{tot}}.
\end{align}

\begin{align}
    \omega^{0}  = \frac{\rho_{\text{tot}}^V}{g_{\omega}\sigma^2}.
\end{align}
and
\begin{align}
    \rho^{03} = \frac{g_{\rho}}{m^{2}_{\rho}}\sum_{f = n,p}I^3_f\rho_f^V + \overline{I}^3_f\overline{\rho}_f^V
\end{align} 
where $I^3_f$ and $\overline{I}^3_f$ are isospin of fermions. The EoM for $\sigma$ is
\begin{align}
&~ \frac{2C_{\sigma}\rho^S}{Y(m_N - g_{\phi}\phi_{0})} 
- \frac{2C_{\sigma}C_{\omega}{\rho^V}^2}{Y^4(m_N - g_{\phi}\phi_{0})^2} -\overline{Y}^2 + \frac{B}{C_{\omega}}\overline{Y}^4 - \frac{C}{C^2_{\omega}}\overline{Y}^6 = 0.
\label{eom:sigma}
\end{align}
\section{GFDI and related expressions}
\label{GFD_Observables}
\noindent One can write numerically tractable forms of all the physical quantities in terms of GFDIs, $F_j(\eta_f, \alpha_f)$ and $G_j(\overline{\eta}_f, \alpha_f)$, for particles and antiparticles respectively, compactified in the following notation
\begin{align}
\mathcal{F}_{j}(\eta, \alpha) = 
\int_{0}^{\infty} dx\,
\frac{x^j \sqrt{1+\frac{x}{2}\alpha}}
     {1+e^{x-\eta}},
\nonumber\\
\end{align}
where $F_j(\eta_f, \alpha_f) = \mathcal{F}_j(\eta_f, \alpha_f)$ and $G_j(\overline{\eta}_f, \alpha_f) = \mathcal{F}_j(-\overline{\eta}_f, \alpha_f)$, $\alpha_f = \frac{T}{m^\star_f}$, $\eta_f = \frac{\mu_f^\star - m_f^\star}{T}$ and $\overline{\eta}_f = \frac{\mu_f^\star + m_f^\star}{T}$. In terms of $\nu^{(i)}_f = \{\eta^{(i)}_{f}, \alpha_f\}$ the scalar densities ($\rho_f^{s}$ and $\overline{\rho}_f^{s}$) are given by
\begin{equation}
\rho_f^{s}(\nu^{(i)}_f)
=
\frac{\gamma_f}{2\pi^2}\sqrt{2}\,\alpha_f^{3/2}m_f^{*3}
\mathcal{F}_{1/2}(\nu^{(i)}_f).
\end{equation}
where $\eta_f^{(+)} = \eta_f$ for particles, $\eta_f^{(-)} = -\overline{\eta}_f$ for antiparticles,
$\rho_f^{s}(\nu^{(+)}_f) = \rho_f^{s}$ and $\rho_f^{s}(\nu^{(-)}_f) = \overline{\rho}_f^{s}$. The vector densities is given by
\begin{align}
\rho_f^V(\nu^{(i)}_f)
&=
\frac{\gamma_f}{2\pi^2}m_f^{\star 3}
\bigg[
\sqrt{2\alpha_f^5}\,
\mathcal{F}_{3/2}(\nu^{(i)}_f)
+
\sqrt{2\alpha_f^3}\,
\mathcal{F}_{1/2}(\nu^{(i)}_f)
\bigg].
\end{align}
The kinetic contributions to pressure due to particles ($\mathcal{P}_f$) and antiparticles ($\overline{\mathcal{P}}_f$) is given by
\begin{align}
\label{Pf}
\mathcal{P}_f(\nu^{(i)}_f)
&=
\frac{\sqrt{2}\,\gamma_f\,\alpha_f^{5/2}m_f^{\star 4}}{3\pi^2}
\left[
\mathcal{F}_{\frac{3}{2}}(\nu^{(i)}_f)
+
\frac{\alpha_f}{2}
\mathcal{F}_{\frac{5}{2}}(\nu^{(i)}_f)
\right].
\end{align}
Similar contribution to the energy density ($\mathcal{E}_f$ and $\overline{\mathcal{E}}_f$) can be written as 
\begin{align}
\mathcal{E}_f(\nu^{(i)}_f) &=
\frac{\sqrt{2}\gamma_f \alpha_f^{5/2} m_f^{\star 4}}{\pi^2}
\Big[
\mathcal{F}_{3/2}(\nu^{(i)}_f)
+ \frac{1}{2\alpha_f} \mathcal{F}_{1/2}(\nu^{(i)}_f)
\nonumber\\
&\phantom{{}=\frac{\sqrt{2}\gamma_f \alpha_f^{5/2} m_f^{\star 4}}{\pi^2}\Big[\mathcal{F}_{3/2}(\nu^{(i)}_f)}
+ \frac{\alpha_f}{2} \mathcal{F}_{5/2}(\nu^{(i)}_f)
\Big]
\label{Ef}
\end{align}
Lastly, the expression of entropy ($S_f$ and $\overline{S}_f$) is given by

\begin{align}
S_f(\nu^{(i)}_f)
&=
\frac{\gamma_f m_f^{\star 3}}{2\pi^2}
\bigg[
\sqrt{2\alpha_f^5}\mathcal{K}_{3/2}(\nu^{(i)}_f)
+
\sqrt{2\alpha_f^3}\mathcal{K}_{1/2}(\nu^{(i)}_f)
\bigg].
\end{align}
where the function $\mathcal{K}_{j}(\nu = \{\eta, \alpha\})$ is given by 
\begin{align}
\mathcal{K}_{j}(\nu)
&=
\int_{0}^{\infty} dx~x^j
\sqrt{1+\frac{x}{2}\alpha}
\bigg[
\ln(1+e^{\eta - x})
+
\frac{x-\eta}{1+e^{x-\eta}}
\bigg].
\end{align}

\def\bibfont{\small}
\bibliography{First_Draft_Ref}{}

\end{document}